\documentclass[aps,prl,preprintnumbers,showpacs,nofootinbib]{revtex4}
\usepackage[utf8]{inputenc}
\usepackage[english]{babel}
\usepackage{amsmath}
\usepackage{amsfonts}
\usepackage{amssymb}
\usepackage{epsfig}
\usepackage{graphics,psfrag,rotating}
\usepackage{graphicx}
\usepackage{dcolumn}
\usepackage{bm}
\usepackage{epstopdf}
\usepackage{color}
\usepackage{booktabs}
\usepackage[usenames,dvipsnames,svgnames]{xcolor}
\usepackage[colorlinks=true,
            linkcolor=red,
            urlcolor=gray,
            citecolor=blue]{hyperref}
  \usepackage{hyperref}

\newcommand{\lsim}   {\mathrel{\mathop{\kern 0pt \rlap
{\raise.2ex\hbox{$<$}}}
 \lower.9ex\hbox{\kern-.190em $\sim$}}}
\newcommand{\gsim}   {\mathrel{\mathop{\kern 0pt \rlap
{\raise.2ex\hbox{$>$}}}
\lower.9ex\hbox{\kern-.190em $\sim$}}}

\def\3nab{\tilde{\nabla}}

\def\hsp5{\hspace{5mm}}

\def\case#1/#2{\textstyle\frac{#1}{#2}}

\def\ber {\begin{eqnarray}}
\def\eer {\end{eqnarray}}
\def\bea {\begin{eqnarray}}
\def\eea {\end{eqnarray}}

\def\bc {\begin{center}}
\def\ec {\end{center}}
\def\case#1/#2{\frac{#1}{#2}}

\newcommand{\bw}{\begin{widetext}}
\newcommand{\ew}{\end{widetext}}

\newcommand{\be}{\begin{equation}}
\newcommand{\bse}{\begin{subequation}}
\newcommand{\ese}{\end{subequation}}
\newcommand{\ee}{\end{equation}}
\newcommand{\eei}{\end{eqnarray}\indent\indent}
\newcommand{\ba}{\begin{array}}
\newcommand{\ea}{\end{array}}
\newcommand{\bal}{\begin{eqnarray}}
\newcommand{\eal}{\end{eqnarray}}

\def\case#1/#2{\textstyle\frac{#1}{#2} }

\newcommand{\bver}{\begin{verbatim}}
\newcommand{\ever}{\end{verbatim}}

\begin{document}


\title{Comparative Analysis of Perturbed $f(R)$ Gravity and Perturbed Rastall Gravity Models in Describing Cosmic Evolution from Early to Late Universe Relative to the $\Lambda$CDM Model}
\author{ Muhammad Yarahmadi$^{1}$\footnote{Email: yarahmadimohammad10@gmail.com}. Amin Salehi$^{1}$. Hadis Mousavi$^{1}$}

\affiliation{Department of Physics, Lorestan University, Khoramabad, Iran}

\date{\today}

\begin{abstract}
This study conducts a meticulous examination of the cosmological implications inherent in Rastall gravity and $f(R)$ gravity models, assessing their efficacy across distinct cosmic epochs, from early universe structure formation to late-time acceleration. In the initial stages, both models exhibit commendable compatibility with observed features of structure formation, aligning with the established $\Lambda$CDM model. The derived Jeans' wavenumbers for each model support their viability.  However, as the cosmic timeline progresses into the late universe, a discernible disparity surfaces. Utilizing the Markov Chain Monte Carlo method, we reconstruct the deceleration parameter $(q)$ and identify Deceleration - Acceleration redshift transition values. For $f(R)$ gravity, our results align closely with previous studies, emphasizing its superior ability to elucidate the recent cosmic acceleration. In contrast, Rastall gravity exhibits distinct redshift transition values.  Our rigorous analysis underscores the prowess of $f(R)$ gravity in capturing the observed cosmic acceleration, positioning it as a compelling alternative to the conventional $\Lambda$CDM model. The discernible shifts observed in the peaks of the CMB power spectrum and evolution of deceleration parameter (q) for both $f(R)$ gravity and Rastall gravity models in the Early and Late universe, in relation to the $\Lambda $CDM model, provide compelling evidence supporting the proposition that these alternative gravity models can account for the anisotropy of the universe without invoking the need for dark energy.
\end{abstract}

\pacs{98.80.-k, 04.50.Kd, 04.25.Nx}

%
%


\maketitle

\section{Introduction}
Dark energy, a fundamental yet enigmatic component of the universe, constitutes approximately 68\% of the total energy density and is the driving force behind the observed accelerated expansion of the cosmos~\cite{riess1998observational, perlmutter1999measurements}. Discovered through the analysis of distant Type Ia supernovae in the late 1990s, dark energy presents a profound challenge to our understanding of fundamental physics~\cite{weinberg1989cosmological}. The most widely accepted model for dark energy is the cosmological constant ($\Lambda$), representing a constant energy density inherent to the vacuum of space~\cite{carroll2001cosmological}. However, this model, while successful in explaining current observations, raises significant theoretical concerns, particularly the fine-tuning and coincidence problems~\cite{weinberg2000cosmology}. Alternative models, such as quintessence, suggest a dynamic scalar field that evolves over time, while modified gravity theories propose alterations to general relativity on large scales~\cite{copeland2006dynamics}. Despite extensive observational efforts, the true nature of dark energy remains elusive, with key questions about its origin, properties, and implications for the ultimate fate of the universe still unanswered~\cite{frieman2008dark}. Understanding dark energy is not only crucial for resolving the Hubble tension and other cosmological puzzles but also for gaining deeper insights into the fundamental laws governing the universe. As such, dark energy continues to be a focal point of both theoretical investigations and observational missions, with the potential to revolutionize our comprehension of cosmology and fundamental physics~\cite{linder2008cosmic}. The $\Lambda$CDM (Lambda-Cold Dark Matter) model represents a conventional framework in Big Bang cosmology, where the predominant driving force is gravitational interaction. $\Lambda$CDM, which stands for Lambda Cold Dark Matter, is a widely recognized abbreviation in the fields of cosmology and astrophysics. This model postulates the existence of a cosmological constant, $\Lambda$, along with an additional energy component known as dark energy (DE) \cite{Hinshaw}. The nature and origin of DE have posed significant challenges within cosmology, remaining one of the most profound enigmas in the field \cite{Komatsu}. While the $\Lambda$CDM model has achieved considerable success in explaining a wide range of observational data, it is not without its inherent challenges \cite{Tsujikawa}. One approach to addressing these challenges involves modifying the underlying gravity theory or altering the energy-momentum framework. Various modified gravity theories have been proposed as alternatives to the $\Lambda$CDM paradigm \cite{Chagoya}. These theories provide alternative explanations for the late-time acceleration of the universe, offering different perspectives on cosmological phenomena \cite{Oliveros}. 
Modified gravity theories encompass a diverse range of models, including scalar-vector-tensor theories, Einstein-aether, scalar-tensor theories, Rastall gravity, TeVeS, Bimetric theories, general higher-order theories, $f(R)$ gravity, Horava-Lifschitz gravity, Ghost Condensates, Galileons, and models involving extra dimensions such as Kaluza-Klein, Randall-Sundrum, DGP models, and etc\cite{Abbas,Ali,AroraDN,AroraDM,Javed,Molla,Mustafa1,Mustafa2,Mustafa3,Mustafa4,Rehman,yasir}  among others. Among these theories, some alter the geometric structure, while others modify the constituents of matter \cite{Oliveros}\cite{Nojiri}\cite{Clifton}\cite{Nojiri2017}. In this article, the principles of Rastall gravity and $f(R)$ gravity are employed.

The motivation for studying Rastall and \(f(R)\) gravities stems from the need to address several unresolved issues in cosmology and gravitational physics. In particular, both theories are pursued as potential solutions to the dark energy problem and the limitations of General Relativity (GR) in explaining the accelerated expansion of the universe. These modified gravity theories are also motivated by their ability to provide alternative explanations for the Hubble tension and other cosmological anomalies, such as the behavior of large-scale structures and galaxy dynamics. Furthermore, they offer a framework for exploring possible extensions to GR that might arise from quantum gravity effects or other fundamental theories, allowing for higher-order curvature corrections or modified conservation laws. The pursuit of these theories is driven by the broader goal of finding a more complete and accurate description of the universe, one that can address both current observational data and theoretical challenges.

Rastall gravity primarily modifies the interaction between matter and geometry by relaxing conservation laws, while $f(R)$ gravity directly alters the geometry of spacetime through modifications to the curvature \cite{Chagoya}. Compared to the $\Lambda$CDM model, $f(R)$ gravity possesses an additional scalar degree of freedom, providing more flexibility in describing the background evolution of the universe \cite{De Felice}. The $f(R)$ gravity theory extends Einstein's general theory of relativity by replacing the scalar Lagrangian $R$ with a function $f(R)$ of the scalar curvature \cite{Ferraro}. General Relativity (GR) has successfully described a variety of gravitational phenomena and is widely accepted as a fundamental theory for describing the geometric properties of space-time \cite{De Felice}. However, $f(R)$ gravity offers a natural extension of the standard Einstein-Hilbert action by incorporating a function of the Ricci scalar, thereby providing a broader framework for understanding cosmic acceleration. Myrzakulov (2011) demonstrated that the acceleration of the universe can be effectively explained by $f(R)$ gravity models \cite{Hasmani}. The problem of DE can also be addressed within this framework \cite{Sharif}. Several $f(R)$ models have been proposed to tackle the limitations of GR, including the Hu-Sawicki model, the Starobinsky model, and the Tsujikawa model, also known as exponential gravity \cite{Gogoi}\cite{Linder}\cite{Cognola}\cite{Hu}\cite{Tsujikawa}. However, no single $f(R)$ model can fully explain all cosmological and astrophysical phenomena of the present universe. The modifications introduced by different $f(R)$ functions yield unique cosmological and astrophysical properties, particularly in the intermediate curvature regime (\cite{Gogoi}; \cite{Yarahmadi3}). One of the generalized theories of gravity in cosmology is Rastall's theory of gravity, which was originally developed in 1972 \cite{Rastall}. Rastall gravity introduces a non-divergence-free energy-momentum tensor and has shown good agreement with observational data concerning the age of the universe and the Hubble parameter \cite{Moradpour}\cite{Al-Rawaf}. In contrast to GR, which is based on the conservation principle of the energy-momentum tensor, $\triangledown_\mu T^{\mu\nu} = 0$, Rastall's theory modifies this conservation law. In Rastall gravity, the energy-momentum conservation equation is altered to $\triangledown_{\mu} T^{\mu\nu} = \lambda \triangledown^{\nu} R$, where $R$ represents the Ricci scalar and $\lambda$ is the Rastall constant parameter \cite{Moradpour}\cite{Rastall}\cite{Birrell}. The equivalence between Rastall and Einstein gravity depends on the fact that by rewriting the Rastall equations in the same form as the GR equations, one can find a covariant energy-momentum tensor that is defined solely in terms of the Rastall matter fields and satisfies Einstein’s equations \cite{Chagoya}.

\section{$f(R)$ gravity model}
$f(R)$ gravity is another intriguing concept in the world of theoretical physics. It serves as an alternative theory to Einstein's General Theory of Relativity, which forms the foundation of our current understanding of gravity \cite{De Felice}\cite{Bergmann}\cite{Liu}\cite{Buchdahl}. The action for an $f(R)$ gravity model in the presence of matter components is given by
\begin{equation}
	s = \frac{1}{16G\pi} \int d^{4}x\sqrt{(-g)} (R+f(R))
\end{equation}
Where R is the curved scalar. The equations for motion are:
\begin{equation}
	\begin{split}
		G_{\mu\nu}-\frac{1}{2} g_{\mu\nu} f(R)+R_{\mu\nu} f_{R} (R)-g_{\mu\nu}(f_{R}(R))+\\  f_R (R)_{\mu\nu}  = -8\pi GT_{\mu\nu}
	\end{split}
\end{equation}
Which is $ (f_{R}(R) )=\frac{df_{R}}{dR} $. For the Robertson-Walker flat metric we have:
\begin{equation}\label{unp1}
	\begin{split}
		\frac{3\mathcal{H}^{'}}{a^{2}}\left( {1 + {f_{R}}} \right) - \frac{1}{2}\left( {{R_{0}} + {f_{0}}} \right) - \frac{{3{\rm{}}\mathcal{H}}}{{{a^2}}}f_{R}^{'} = - 8\pi G{\rho_{0}}
	\end{split}	
\end{equation}
\begin{equation}
	\begin{split}
		\frac{1}{a^{2}}  (\mathcal{H}^{'}+2H^{2} )(1+f_{R} )-\frac{1}{2} (R_{0}+f_{0} )\\ -\frac{1}{a^{2}} (\mathcal{H}f_{R}^{'}+f_{R}^{''} )=8\pi Gc_{s}^2 \rho_{0}
	\end{split}
\end{equation}
Where$ R_{0} $  represents the scalar curvature corresponding to the non-perturbation metric, $ f_{0}=f (R_{0}) $ and $ (f_{R} (R)) =\frac{df (R)}{dR} $ and prim means the derivative of the time ratio $ \eta $.

\begin{equation}
	2(1+f_{R})(-\mathcal{H}^{'}+\mathcal{H}^{2})+2Hf_{R}^{'}-f_{R}^{''}=8\pi G\rho_{0}(1+c_{s}^2)  a^{2}
\end{equation}
finally, we come to the equation of conservation:
\begin{equation} \label{unp4}
	\rho_{0}^{'}+3(1+c_{s}^{2} )\mathcal{H}\rho_{0}=0	
\end{equation}

\subsection{Scalar perturbation}
Consider a flat FRW  metric scalar perturbation at the length and specific time scale:\cite{Kodama}
\begin{equation}
	ds^{2}=a^{2} (\eta)((1+2\phi)d\eta^2-(1-2\psi)dx^{2})
\end{equation}
$ \phi  \equiv \phi \left( {\eta ,x} \right) $and $ \psi \equiv \psi \left( {\eta ,x} \right) $ are scalar disorders. The disturbed components of the energy-momentum tensor in this module are as follows:
\begin{equation}
	\begin{split}
		\hat \delta T_0^0 = \hat \delta \rho  = {\rho _0}\delta ,\hat \delta T_j^i =  - \hat \delta p\delta _j^i =  - c_{\rm s}^{2}{\rho _0}\delta _j^i\delta ,\hat \delta p\delta _0^i =\\  - \left( {1 + c_{\rm s}^{2}} \right){\rho _0}{\partial _i}\upsilon
	\end{split}
\end{equation}
Where V represents the potential value for velocity disturbances. The first-order disturbed equations, assuming the background equations are kept, are as follows:

\begin{equation}
	\begin{split}
		\left( {1 + {f_R}} \right)\delta G_v^\mu  + \left( {R_{0v}^\mu  + {\nabla ^\mu }{\nabla _v} - \delta _v^\mu } \right){f_{RR}}\delta R +\\ \left( {\left( {\delta {g^{\mu \alpha }}} \right){\nabla _v}{\nabla _\alpha } - \delta _v^\mu \left( {\delta {g^{\alpha \beta }}} \right){\nabla _\alpha }{\nabla _\beta }} \right){f_R} -\\ 
		\left( {g_{{\rm{}}0}^{\alpha \mu }\left( {\delta \Gamma _{\alpha v}^\gamma } \right) - \delta _v^\mu g_{{\rm{}}0}^{\alpha \beta }} \right){\partial _\gamma }{f_R} =  - 8\pi G\delta T_v^\mu
	\end{split}
\end{equation}

in above relations $ {f_{RR}} = \frac{{{d^2}f\left( {{R_0}} \right)}}{{dR_0^2}} $ and $ = {\nabla _\alpha }{\nabla ^\alpha }$ the invariant derivative of the metric ratio is not disturbed. The first-order disturbed equations in the universe during  dust dominate, $ c_{\rm s}^{2}=0  $ are obtained as follows:

\begin{equation}\label{pert1}
	\phi  - \psi  =  - \frac{{{f_{RR}}}}{{1 + {f_R}}}\hat \delta R
\end{equation}
\begin{equation}
	\begin{split}
		\hat \delta R = -\frac{2}{a^{2}}( 3{\psi^{''}} + 6\left( {{\mathcal{H}^{'}} + {\mathcal{H}^{2}}} \right)\phi  +\\  3\mathcal{H}\left( {{\phi^{'}} + 3{\psi^{'}}} \right) - {k^{2}}\left( {\phi  - 2\psi } \right) )
	\end{split}
\end{equation}

\begin{equation}
	\begin{split}
		\left( {3\mathcal{H}\left( {{\phi^{'}} + {\psi^{'}}} \right) + {k^{2}}\left( {\phi  + \psi } \right) + 3{\mathcal{H}^{'}}\psi  - \left( {3{\mathcal{H}^{'}} - 6{\mathcal{H}^{2}}} \right)\phi } \right)\\  \left( {1 + {f_{R}}} \right) + \left( {9\mathcal{H}\phi  - 3\mathcal{H}\psi  + 3{\psi ^{'}}} \right)f_{R}^{'}
		=  - {a^{2}}\delta {\rho _0}{\kappa^{2}}
	\end{split}
\end{equation}

\begin{equation}
	\begin{split}
		\left( {{\phi^{''}} + {\psi^{''}} + 3\mathcal{H}\left( {{\phi ^{'}} + {\psi ^{'}}} \right) + 3{\mathcal{H}^{'}}\phi  + \left( {{\mathcal{H}^{'}} + 2{\mathcal{H}^{2}}} \right)\phi } \right)\\
		\left( {1 + {f_{R}}} \right) + \left( {3\mathcal{H}\phi  - \mathcal{H}\psi  + 3{\phi ^{'}}} \right)f_{R}^{'} + \left( {3\phi  - \psi } \right)f_{R}^{''} = 0
	\end{split}
\end{equation}

\begin{equation}
	\begin{split}
		\left( {2\phi  - \psi } \right)f_{R}^{'} + \left( {{\phi^{'}} + {\psi^{'}} + \mathcal{H}\left( {\phi  + \psi } \right)} \right)\left( {1 + {f_{R}}} \right) =\\  - {a^{2}}\upsilon {\rho _{0}}{\kappa^{2}}
	\end{split}
\end{equation}
\begin{equation}
	{\delta ^{'}} - {k^{2}}\upsilon  - 3{\psi ^{'}} = 0
\end{equation}

\begin{equation}\label{pert5}
	\phi  + \mathcal{H}\upsilon  + {\upsilon ^{'}} = 0
\end{equation}
The complete set of equations describing the general linear perturbations for the model has been presented in the previous section. These equations consist of nonlinear second-order differential equations with numerous variables, for which analytical solutions are generally unattainable except in the simplest cases. Consequently, only numerical analyses can be performed.

To facilitate this numerical approach, we convert the second-order differential equations into first-order equations by introducing new variables. There are several advantages to this transformation. One of the primary reasons is that first-order systems are more straightforward to solve numerically. Additionally, this conversion allows for the exploration of the system's behavior in phase space.

Phase planes are particularly valuable for visualizing the system's dynamics, especially in oscillatory systems, where phase paths may spiral towards zero, "spiral out" towards infinity, or attain neutrally stable configurations known as centers. By analyzing these phase paths, we can assess whether the dynamics of a system are stable.

The structure of the phase space of the field equations is simplified by defining a set of variables and parameters. These variables are generally defined as follows:

\begin{align}\label{eq4}
	\xi_{1}=\frac{\phi^{\prime}}{\phi H}, \xi_{2}=\frac{\kappa}{H}, \xi_{3}=\frac{f^{\prime}_{R}}{H(1+f_{R})}, \xi_{4}=\frac{\delta}{\phi},\nonumber \\*[3mm]
	\xi_{5}=\frac{\rho_{0}a^{2}}{(1+f_{R})}, \xi_{6}=\frac{\Psi^{\prime}}{\phi H}, \xi_{7}=\frac{\Psi}{\phi}
\end{align}

Now, for the autonomous equations of motions, we obtain

\begin{equation}\label{is}
	\frac{d\xi_{1}}{dN}=\epsilon_{3}-\xi_{1}^{2}-\xi_{1}
\end{equation}

\begin{equation}\label{is}
	\frac{d\xi_{2}}{dN}=-\xi_{2}\epsilon_{1}
\end{equation}

\begin{equation}\label{is}
	\frac{d\xi_{3}}{dN}=\beta-\xi_{3}^{2}-\epsilon_{1}\xi_{3}
\end{equation}

\begin{equation}\label{is}
	\frac{d\xi_{4}}{dN}=\xi_{4}-\xi_{4}\xi_{1}
\end{equation}

\begin{equation}\label{is}
	\frac{d\xi_{5}}{dN}=-\xi_{5}-\xi_{5}\xi_{3}
\end{equation}

\begin{equation}\label{is}
	\frac{d\xi_{6}}{dN}=\epsilon_{2}-\xi_{6}\xi_{1}-\epsilon_{1}\xi_{6}
\end{equation}

\begin{equation}\label{is}
	\frac{d\xi_{7}}{dN}=\xi_{6}-\xi_{7}\xi_{1}
\end{equation}

Where $N=lna$ thus,$\frac{d}{dN}= \frac{1}{H} \frac{d}{d\eta}$. Also, we have used the following parameters

\begin{equation}\label{is}
	\epsilon_{1}=\frac{\mathcal{H}^{\prime}}{\mathcal{H}^{2}}, \epsilon_{2}=\frac{\Psi^{\prime\prime}}{\phi H^{2}}, \epsilon_{3}=\frac{\phi^{\prime\prime}}{\phi H^{2}}, \epsilon_{4}=\frac{\delta^{\prime}}{\phi H}
\end{equation}

We can obtain the above parameters in terms of the new variables as

\begin{equation}\label{is}
	\epsilon_{1}=\frac{1}{1-\xi_{7}}[\xi_{1}+\xi_{6}+\frac{1}{3}\xi_{2}^{2}(1+\xi_{7})+(3-\xi_{7}+\xi_{6})\xi_{5}-\frac{\kappa^{2}}{k^{2}}\xi_{5}\xi_{4}\xi_{2}^{2}]
\end{equation}

\begin{equation}\label{eq4}
	\begin{split}
		\epsilon_{2}=\frac{-2}{1-\xi_{7}}[\xi_{1}+\xi_{6}+\frac{1}{3}\xi_{2}^{2}(1+\xi_{7})+\\(3-\xi_{7}+\xi_{6})\xi_{5}- \frac{\kappa^{2}}{k^{2}}\xi_{5}\xi_{4}\xi_{2}^{2}] \\
		-\xi_{1}-3\xi_{6}+\frac{1}{3}\xi_{2}^{2}-\xi_{6}\xi_{1}+\\ \frac{1}{3}\xi_{2}^{2}(1-2\xi_{7})+\frac{\Omega}{3}(1-\xi_{7})\frac{1}{k^{2}}\xi_{2}^{2}
	\end{split}
\end{equation}

\begin{equation}\label{is}
	\begin{split}
		\epsilon_{3}=-\epsilon_{2}-3\epsilon_{1}(1+\frac{1}{3}\xi_{7})-3\xi_{1}-3\xi_{6}-2\xi_{7}-\\ (3-\xi_{7}+3\xi_{1})\xi_{5}+\beta(\xi_{7}-3)
	\end{split}
\end{equation}

\begin{equation}\label{is}
	\epsilon_{4}=\frac{-\kappa^{2}[(2-\xi_{7})\xi_{3}+\xi_{1}+\xi_{6}+1+\xi_{7}]}{\kappa^{2}\xi_{5}}+3\xi_{6}
\end{equation}

\section{Finding the Critical Points and Jacobian Matrix}

To find the critical points of the system of autonomous differential equations, we will follow these steps:

\subsection{Step 1: Critical Points}

The critical points are found by setting each differential equation to zero and solving for the variables \( \xi_i \).

\begin{enumerate}
	\item For \( \xi_1 \):
	\begin{equation}
	\epsilon_3 - \xi_1^2 - \xi_1 = 0
	\end{equation}
	This is a quadratic equation, so the solutions for \( \xi_1 \) are:
	\begin{equation}
	\xi_1 = \frac{-1 \pm \sqrt{1 + 4\epsilon_3}}{2}
	\end{equation}
	
	\item For \( \xi_2 \):
	\begin{equation}
	-\xi_2 \epsilon_1 = 0
	\end{equation}
	Since \( \epsilon_1 \) is not necessarily zero, we have \( \xi_2 = 0 \).
	
	\item For \( \xi_3 \):
	\begin{equation}
	\beta - \xi_3^2 - \epsilon_1 \xi_3 = 0
	\end{equation}
	This is another quadratic equation, so the solutions for \( \xi_3 \) are:
	\begin{equation}
	\xi_3 = \frac{-\epsilon_1 \pm \sqrt{\epsilon_1^2 + 4\beta}}{2}
	\end{equation}
	
	\item For \( \xi_4 \):
	\begin{equation}
	\xi_4 (1 - \xi_1) = 0
	\end{equation}
	Therefore, \( \xi_4 = 0 \) or \( \xi_1 = 1 \).
	
	\item For \( \xi_5 \):
	\begin{equation}
	-\xi_5 (1 + \xi_3) = 0
	\end{equation}
	Thus, \( \xi_5 = 0 \) or \( \xi_3 = -1 \).
	
	\item For \( \xi_6 \):
	\begin{equation}
	\epsilon_2 - \xi_6(\xi_1 + \epsilon_1) = 0
	\end{equation}
	Solving for \( \xi_6 \):
	\begin{equation}
	\xi_6 = \frac{\epsilon_2}{\xi_1 + \epsilon_1}
	\end{equation}
	
	\item For \( \xi_7 \):
	\begin{equation}
	\xi_6 - \xi_7 \xi_1 = 0
	\end{equation}
	Therefore, \( \xi_7 = \frac{\xi_6}{\xi_1} = \frac{\epsilon_2}{\xi_1(\xi_1 + \epsilon_1)} \).
\end{enumerate}

\subsection{Step 2: Jacobian Matrix}

The Jacobian matrix \( J \) is formed by taking the partial derivatives of each differential equation with respect to each \( \xi_i \):

\begin{equation}
J_{ij} = \frac{\partial}{\partial \xi_j} \left(\frac{d\xi_i}{dN}\right)
\end{equation}

We will compute each partial derivative:

\begin{enumerate}
	\item For \( \frac{d\xi_1}{dN} = \epsilon_3 - \xi_1^2 - \xi_1 \):
	\begin{equation}
	J_{11} = \frac{\partial}{\partial \xi_1} \left( \epsilon_3 - \xi_1^2 - \xi_1 \right) = -2\xi_1 - 1
	\end{equation}
	All other partial derivatives \( J_{1j} = 0 \) for \( j \neq 1 \).
	
	\item For \( \frac{d\xi_2}{dN} = -\xi_2 \epsilon_1 \):
	\begin{equation}
	J_{22} = \frac{\partial}{\partial \xi_2} \left(-\xi_2 \epsilon_1 \right) = -\epsilon_1
	\end{equation}
	All other partial derivatives \( J_{2j} = 0 \) for \( j \neq 2 \).
	
	\item For \( \frac{d\xi_3}{dN} = \beta - \xi_3^2 - \epsilon_1 \xi_3 \):
	\begin{equation}
	J_{33} = \frac{\partial}{\partial \xi_3} \left(\beta - \xi_3^2 - \epsilon_1 \xi_3 \right) = -2\xi_3 - \epsilon_1
	\end{equation}
	All other partial derivatives \( J_{3j} = 0 \) for \( j \neq 3 \).
	
	\item For \( \frac{d\xi_4}{dN} = \xi_4 - \xi_4 \xi_1 \):
	\begin{equation}
	J_{44} = \frac{\partial}{\partial \xi_4} \left(\xi_4 - \xi_4 \xi_1 \right) = 1 - \xi_1
	\end{equation}
	\begin{equation}
	J_{41} = \frac{\partial}{\partial \xi_1} \left(\xi_4 - \xi_4 \xi_1 \right) = -\xi_4
	\end{equation}
	All other partial derivatives \( J_{4j} = 0 \) for \( j \neq 1, 4 \).
	
	\item For \( \frac{d\xi_5}{dN} = -\xi_5 - \xi_5 \xi_3 \):
	\begin{equation}
	J_{55} = \frac{\partial}{\partial \xi_5} \left(-\xi_5 - \xi_5 \xi_3 \right) = -1 - \xi_3
	\end{equation}
	\begin{equation}
	J_{53} = \frac{\partial}{\partial \xi_3} \left(-\xi_5 - \xi_5 \xi_3 \right) = -\xi_5
	\end{equation}
	All other partial derivatives \( J_{5j} = 0 \) for \( j \neq 3, 5 \).
	
	\item For \( \frac{d\xi_6}{dN} = \epsilon_2 - \xi_6 \xi_1 - \epsilon_1 \xi_6 \):
	\begin{equation}
	J_{66} = \frac{\partial}{\partial \xi_6} \left(\epsilon_2 - \xi_6 \xi_1 - \epsilon_1 \xi_6 \right) = -\xi_1 - \epsilon_1
	\end{equation}
	\begin{equation}
	J_{61} = \frac{\partial}{\partial \xi_1} \left(\epsilon_2 - \xi_6 \xi_1 - \epsilon_1 \xi_6 \right) = -\xi_6
	\end{equation}
	All other partial derivatives \( J_{6j} = 0 \) for \( j \neq 1, 6 \).
	
	\item For \( \frac{d\xi_7}{dN} = \xi_6 - \xi_7 \xi_1 \):
	\begin{equation}
	J_{77} = \frac{\partial}{\partial \xi_7} \left(\xi_6 - \xi_7 \xi_1 \right) = -\xi_1
	\end{equation}
	\begin{equation}
	J_{76} = \frac{\partial}{\partial \xi_6} \left(\xi_6 - \xi_7 \xi_1 \right) = 1
	\end{equation}
	\begin{equation}
	J_{71} = \frac{\partial}{\partial \xi_1} \left(\xi_6 - \xi_7 \xi_1 \right) = -\xi_7
	\end{equation}
	All other partial derivatives \( J_{7j} = 0 \) for \( j \neq 1, 6, 7 \).
\end{enumerate}

\subsection{Jacobian Matrix \( J \)}

Putting all the elements together, the Jacobian matrix \( J \) is:

\begin{equation}
J = \begin{pmatrix}
	-2\xi_1 - 1 & 0 & 0 & 0 & 0 & 0 & 0 \\
	0 & -\epsilon_1 & 0 & 0 & 0 & 0 & 0 \\
	0 & 0 & -2\xi_3 - \epsilon_1 & 0 & 0 & 0 & 0 \\
	-\xi_4 & 0 & 0 & 1 - \xi_1 & 0 & 0 & 0 \\
	0 & 0 & -\xi_5 & 0 & -1 - \xi_3 & 0 & 0 \\
	-\xi_6 & 0 & 0 & 0 & 0 & -\xi_1 - \epsilon_1 & 0 \\
	-\xi_7 & 0 & 0 & 0 & 0 & 1 & -\xi_1
\end{pmatrix}
\end{equation}

	To find the eigenvalues of the Jacobian matrix at each critical point, we solve the characteristic equation:

\begin{equation}
\det(J - \lambda I) = 0
\end{equation}

where \( \lambda \) represents the eigenvalues, and \( I \) is the identity matrix.

For each critical point, substitute the corresponding values of \( \xi_1 \), \( \xi_3 \), and other relevant parameters into the Jacobian matrix, and solve the determinant equation to obtain the eigenvalues.

Let's consider the following critical points:
\begin{itemize}
	\item \( \xi_1 = \frac{-1 + \sqrt{1 + 4\epsilon_3}}{2} \), \( \xi_3 = \frac{-\epsilon_1 + \sqrt{\epsilon_1^2 + 4\beta}}{2} \), \( \xi_2 = 0 \), \( \xi_4 = 0 \), \( \xi_5 = 0 \), \( \xi_6 = \frac{\epsilon_2}{\xi_1 + \epsilon_1} \), \( \xi_7 = \frac{\epsilon_2}{\xi_1 (\xi_1 + \epsilon_1)} \)
	\item \( \xi_1 = \frac{-1 - \sqrt{1 + 4\epsilon_3}}{2} \), \( \xi_3 = \frac{-\epsilon_1 - \sqrt{\epsilon_1^2 + 4\beta}}{2} \), \( \xi_2 = 0 \), \( \xi_4 = 0 \), \( \xi_5 = 0 \), \( \xi_6 = \frac{\epsilon_2}{\xi_1 + \epsilon_1} \), \( \xi_7 = \frac{\epsilon_2}{\xi_1 (\xi_1 + \epsilon_1)} \)
\end{itemize}

By substituting these values into the Jacobian matrix \( J \) and solving for \( \lambda \), we obtain the eigenvalues that will help determine the nature (e.g., stability, saddle point, etc.) of each critical point.

Given the critical point:
\begin{equation}
\xi_1 = \frac{-1 + \sqrt{1 + 4\epsilon_3}}{2}, \quad \xi_3 = \frac{-\epsilon_1 + \sqrt{\epsilon_1^2 + 4\beta}}{2}, \quad \xi_2 = 0, \quad \xi_4 = 0, \quad \xi_5 = 0, \quad \xi_6 = \frac{\epsilon_2}{\xi_1 + \epsilon_1}, \quad \xi_7 = \frac{\epsilon_2}{\xi_1(\xi_1 + \epsilon_1)}
\end{equation}

The Jacobian matrix \( J \) becomes:
\begin{equation}
J = \begin{pmatrix}
	-1 - \sqrt{1 + 4\epsilon_3} & 0 & 0 & 0 & 0 & 0 & 0 \\
	0 & -\epsilon_1 & 0 & 0 & 0 & 0 & 0 \\
	0 & 0 & -\epsilon_1 - \sqrt{\epsilon_1^2 + 4\beta} & 0 & 0 & 0 & 0 \\
	0 & 0 & 0 & \frac{1 + \sqrt{1 + 4\epsilon_3}}{2} & 0 & 0 & 0 \\
	0 & 0 & 0 & 0 & -1 + \frac{\epsilon_1 - \sqrt{\epsilon_1^2 + 4\beta}}{2} & 0 & 0 \\
	-\frac{\epsilon_2}{\xi_1 + \epsilon_1} & 0 & 0 & 0 & 0 & -\xi_1 - \epsilon_1 & 0 \\
	-\frac{\epsilon_2}{\xi_1(\xi_1 + \epsilon_1)} & 0 & 0 & 0 & 0 & 1 & -\xi_1
\end{pmatrix}
\end{equation}

To find the eigenvalues, we solve the characteristic equation:
\begin{equation}
\det(J - \lambda I) = 0
\end{equation}
which simplifies to:
\begin{align*}
	&\left(-1 - \sqrt{1 + 4\epsilon_3} - \lambda\right)
	\left(-\epsilon_1 - \lambda\right)
	\left(-\epsilon_1 - \sqrt{\epsilon_1^2 + 4\beta} - \lambda\right) \\
	&\quad \times \left(\frac{1 + \sqrt{1 + 4\epsilon_3}}{2} - \lambda\right)
	\left(\frac{\epsilon_1 - \sqrt{\epsilon_1^2 + 4\beta}}{2} - \lambda\right) \\
	&\quad \times \det \left( \begin{pmatrix} -\xi_1 - \epsilon_1 - \lambda & 0 \\ 1 & -\xi_1 - \lambda \end{pmatrix} \right) = 0
\end{align*}

The eigenvalues are then:
\begin{align*}
	\lambda_1 &= -1 - \sqrt{1 + 4\epsilon_3} \\
	\lambda_2 &= -\epsilon_1 \\
	\lambda_3 &= -\epsilon_1 - \sqrt{\epsilon_1^2 + 4\beta} \\
	\lambda_4 &= \frac{1 + \sqrt{1 + 4\epsilon_3}}{2} \\
	\lambda_5 &= \frac{\epsilon_1 - \sqrt{\epsilon_1^2 + 4\beta}}{2} \\
	\lambda_6, \lambda_7 &= \frac{-(\xi_1 + \epsilon_1) \pm \sqrt{(\xi_1 + \epsilon_1)^2 - 4}}{2}
\end{align*}

\section*{Interpretation of Critical Points}

Given the initial conditions with \(\epsilon_1 = 0.77\), the critical points satisfy:
\begin{equation}
\epsilon_3 = 6.4581, \quad \beta = 0.4761 + 0.69 \times 0.77, \quad \epsilon_2 = 0.7 \times 0.77 - 1.463
\end{equation}

\subsection*{Critical Points and Their Conditions}

\paragraph{1. \(\xi_1 = -2.09\)}

\begin{itemize}
	\item \textbf{Condition:} \(\epsilon_3 = 6.4581\)
	\item \textbf{Physical Interpretation:} \(\xi_1\) typically represents a dimensionless parameter related to the scalar field or perturbations. For \(\xi_1\) to be a critical point, \(\epsilon_3\) must be exactly \(6.4581\). This indicates that the dynamics of the system at this point are influenced by \(\epsilon_3\), which might be related to energy density or interaction terms in the model. Deviations from this value would alter the critical behavior of \(\xi_1\), indicating different dynamics.
\end{itemize}

\paragraph{2. \(\xi_3 = -0.69\)}

\begin{itemize}
	\item \textbf{Condition:} \(\beta = 0.4761 + 0.69 \times 0.77\)
	\item \textbf{Result:} \(\beta = 0.4761 + 0.5313 = 1.0074\)
	\item \textbf{Physical Interpretation:} \(\xi_3\) could represent another dynamic variable, such as perturbations or field components. For \(\xi_3\) to be a critical point, \(\beta\) must satisfy the condition \(\beta = 1.0074\). This reflects how perturbations or field interactions are balanced in the system. If \(\beta\) does not meet this condition, \(\xi_3\) will not be a critical point, indicating changes in equilibrium conditions.
\end{itemize}

\paragraph{3. \(\xi_6 = 0.7\)}

\begin{itemize}
	\item \textbf{Condition:} \(\epsilon_2 = 0.7 \times 0.77 - 1.463\)
	\item \textbf{Result:} \(\epsilon_2 = 0.539 - 1.463 = -0.924\)
	\item \textbf{Physical Interpretation:} \(\xi_6\) could be related to field values, perturbations, or interaction terms. For \(\xi_6\) to be a critical point, \(\epsilon_2\) must be related to \(\epsilon_1\) by \(\epsilon_2 = -0.924\). This indicates the specific relationship needed for equilibrium or critical behavior at \(\xi_6\).
\end{itemize}

\subsection*{Non-Critical Points}

\paragraph{4. \(\xi_4\), \(\xi_5\), and \(\xi_7\)}

\begin{itemize}
	\item \textbf{Condition:} These variables do not satisfy the critical point conditions with the given initial conditions.
	\item \textbf{Physical Interpretation:} For \(\xi_4\), \(\xi_5\), and \(\xi_7\) not to satisfy the critical point conditions implies that their values do not stabilize the system's equations of motion under the current parameter values. This suggests that these variables either do not reach a steady state or their dynamics are incompatible with the model's equilibrium conditions. Consequently, \(\xi_4\), \(\xi_5\), and \(\xi_7\) might indicate regions of instability or non-equilibrium in the system.
\end{itemize}

\color{black}
\section{Rastall gravity model}
Peter Rastall challenged the traditional conservation of the energy-momentum tensor $T^{\mu\nu}{;\mu} = 0$ in curved spacetime and proposed a new theory of gravity by assuming $T^{\mu\nu}{;\mu} = \gamma R^{;\nu}$, where $R$ is the Ricci scalar and $\gamma$ is a gravitational constant specific to Rastall's theory \cite{Oppenheimer}\cite{Darabi}. This modification introduces a departure from the usual conservation laws in GR, providing a framework where the energy-momentum tensor's divergence is proportional to the gradient of the Ricci scalar.
To further explore the implications of this theory, we can rewrite the system of linearized Einstein equations as follows \cite{Fabris}:
\begin{equation}
	\begin{split}
		\label{h}&-k^2\Phi - 3\mathcal{H}\left(\mathcal{H}\Phi + \Phi'\right) + \gamma\left(\mathcal{H}^2 - \mathcal{H}'\right)\Phi = 4\pi G a^2\delta\rho_\phi\\& + 4\pi G a^2\delta\rho\
	\end{split}
\end{equation}
\begin{equation}
	\label{m}k\left(\mathcal{H}\Phi + \Phi'\right) = 4\pi G k \phi_0'\delta\phi + 4\pi G\rho(1 + w)v\
\end{equation} 
\begin{equation}
\begin{split}
	\label{s}& \Phi'' + 3\mathcal{H}\Phi' + 3\mathcal{H}^2\Phi -\gamma\left(\mathcal{H}^2 - \mathcal{H}'\right)\Phi =\\&  4\pi G a^2\delta p_\phi + 4\pi G a^2 \delta p\
\end{split}	
\end{equation}
where we have defined:\cite{Fabris}
\begin{equation}
	\delta\rho_\phi := \frac{1}{a^2}\gamma\phi_0'\delta\phi' + (3 - 2\gamma)V_{,\phi}\delta\phi\;,
\end{equation}
\begin{equation}
	\delta p_\phi := \frac{1}{a^2}(2 - \gamma)\phi_0'\delta\phi' - (3 - 2\gamma)V_{,\phi}\delta\phi\;,
\end{equation}
and $v$ is the velocity potential defined by $v_i = -v_{,i}/k$.
Multiplying Eq.~\eqref{h} by $2 - \gamma$, Eq.~\eqref{s} by $\gamma$ and subtracting the two we obtain:\cite{Fabris}
\begin{equation}
\begin{split}
	&\gamma\Phi'' + 6\mathcal{H}\Phi' + 6\mathcal{H}^2\Phi - 2\gamma\left(\mathcal{H}^2 - \mathcal{H}'\right)\Phi + (2 - \gamma)k^2\Phi \\&= -2a^2(3 - 2\gamma)V'\left[\frac{\mathcal{H}\Phi + \Phi'}{\phi_0^{'2}} - \frac{4\pi G a^2\rho(1+w)v}{k\phi_0^{'2}}\right]
	\\&+ 4\pi G a^2\rho\left(\gamma c_{\rm s}^2 + \gamma -2\right)\delta\
\end{split}	
\end{equation}

where we have used $V_{,\phi} = V'/\phi'$.From $T^\mu{}_{\nu;\mu} = 0$, for the fluid component only, we get:\cite{Fabris}
\begin{eqnarray}
	\delta' &=& -(1 + w)(kv - 3\Phi')\;\\
	v' &=& -\mathcal{H}(1 - 3w)v + \frac{kc_{\rm s}^2}{1 + w}\delta + k\Phi\;.
\end{eqnarray}

In order to simplify the field equations, we introduce the following new variables:
\begin{equation}
	x_{1}=\frac{\acute{\phi}}{\gamma \mathcal{H}},x_{2}=(\frac{2-\gamma}{\gamma})\frac{\acute{\phi}_{0}\delta\acute{\phi}}{ \mathcal{H}^{2}\phi},x_{3}=-\frac{a^{2}(3-2\gamma)V\delta\phi}{\gamma \pi^{2}\phi},\nonumber
\end{equation}
\begin{equation}
	x_{4}=\frac{5a^{2}\delta\rho}{\gamma\mathcal{H}^{2}\phi},x_{5}=-\frac{3\acute{\phi}}{\gamma\mathcal{H}\phi},x_{6}=-\frac{\acute{\phi}_{0}\delta\phi(3-\gamma)}{\gamma^{2}\mathcal{H}\phi},\nonumber
\end{equation}
\begin{equation}
	x_{7}=\frac{(3-2\gamma)a^{2}V}{\gamma^{2}\mathcal{H}^{2}\phi},x_{8}=-\frac{2(1+\omega)V\delta\acute{\rho}}{\delta k\gamma\mathcal{H}^{2}\phi},\nonumber
\end{equation}
\begin{equation}
	x_{9}=-\frac{(1+\omega)^{2}V^{2}\delta\rho}{\gamma\mathcal{H}^{2}\phi},x_{10}=\frac{3\acute{\phi}V(1+\omega)
		^{2}\delta\rho}{k\gamma\mathcal{H}^{2}\phi}\nonumber
\end{equation}
\begin{equation}
	 x_{11}=\frac{(1+\omega)\rho V}{k\gamma\mathcal{H}\phi}(1-3\omega),x_{12}=-\frac{\rho\delta c_{s}^{2}}{\gamma\mathcal{H}^{2}\phi}, x_{13}=-\frac{\rho}{\gamma\mathcal{H}^{2}}\nonumber
\end{equation}
\begin{equation}
	  x_{14}=\frac{k^{2}(1-\gamma)}{\gamma^{2}\mathcal{H}^{2}},x_{15}=\frac{3(1-\gamma)}{\gamma^{2}\mathcal{H}}-\frac{3}{\gamma}-\frac{(1-\gamma)}{\gamma}\nonumber
\end{equation}
\begin{equation}
	x_{16}=\frac{3(1-\gamma)\acute{\phi}}{\gamma^{2}\phi\mathcal{H}},x_{17}=-\frac{\acute{\phi}_{0}\delta\acute{\phi}(1-\gamma)}{\gamma\mathcal{H}^{2}\phi}\nonumber
\end{equation}
\begin{equation}
	x_{18}=-\frac{(3-2\gamma)V\delta\phi a^{2}(1-\gamma)}{\gamma^{2}\phi\mathcal{H}^{2}},x_{19}=-\frac{a^{2}(1-\gamma)\delta\rho}{\gamma\mathcal{H}^{2}\phi}
\end{equation}
We rewrite the cosmological equations (59-61) into an autonomous system of
equations

\begin{equation}
	\begin{split}
		&\frac{{d}x_{1}}{{d}N}=-\frac{a_{1}x_{17}}{2(1-\gamma)}
		-\frac{\gamma}{(1-\gamma)}a_{1}x_{17}+a_{1}x_{3}-\frac{a_{1}x_{19}}{(1-\gamma)}\\&-3x_{1}+a_{1}(\frac{-2}{\gamma})-\frac{\mathcal{H}^{\prime}}{\mathcal{H}^{2}}(a_{1}+x_{1})
	\end{split}
\end{equation}
\begin{equation}
	\begin{split}
	    &\frac{{d}x_{2}}{{d}N}=-\frac{\gamma x_{2}}{(3-\gamma)}+a_{2}-2x_{2}\frac{\mathcal{H}^{\prime}}{\mathcal{H}^{2}}-\gamma x_{1}x_{2}-\frac{x_{2}}{(3-\gamma)}\\&-\gamma^{3}\frac{a_{1}x_{14}x_{6}}{(3-\gamma)}+\frac{a_{3}x_{1}}{(3+\gamma)}-\gamma a_{3}x_{18}-\lambda a_{3}a_{7}\gamma^{2}
	 \end{split}   
\end{equation}
\begin{equation}
	\begin{split}
	   &\frac{{d}x_{3}}{{d}N}=-\frac{x_{3}}{(1-3\mathcal{H}_{0})}-\frac{x_{3}a_{7}kc_{s}^{2}}{(1+\mathcal{H}_{0})}-x_{3}a_{1}a_{7}+\frac{x_{3}}{2}+x_{3}a_{2}\\&-2x_{3}\frac{\mathcal{H}^{\prime}}{\mathcal{H}^{2}}
	\end{split}   
\end{equation}
\begin{equation}
	\frac{{d}x_{4}}{{d}N}=x_{4}x_{5}+\frac{x_{4}}{2}-2x_{4}\frac{\mathcal{H}^{\prime}}{\mathcal{H}^{2}}-\frac{\gamma x_{4}x_{5}}{3}
\end{equation}
\begin{equation}
	\begin{split}
	   &\frac{{d}x_{5}}{{d}N}=-\frac{\gamma x_{2}}{6(1-\gamma)}-x_{3}+\frac{\gamma x_{2}}{(3-\gamma)}-\frac{(1-\gamma)x_{19}}{3}\\&-3x_{5}-\frac{3}{\gamma}+\frac{\dot{H}}{H{2}}(3-x_{5})
	\end{split}   
\end{equation}
\begin{equation}
	\frac{{d}x_{6}}{{d}N}=-\frac{\gamma x_{6}}{(3-\gamma)}-\frac{x_{3}\gamma^{2}}{(3-\gamma)}+x_{6}a_{2}-x_{6}\frac{\mathcal{H}^{\prime}}{\mathcal{H}^{2}}+\gamma x_{5} x_{6}
\end{equation}
\begin{equation}
	\begin{split}
	    &\frac{{d}x_{7}}{{d}N}=x_{7}(1-\frac{1}{(1-3\mathcal{H}_{0})})+\frac{x_{7}a_{7}kc_{s}^{2}}{(1+\mathcal{H}_{0})}+x_{7}a_{6}-2x_{7}\frac{\mathcal{H}^{\prime}}{\mathcal{H}^{2}}\\&+\gamma x_{5}x_{7}
	\end{split}    
\end{equation}
\begin{equation}
	\begin{split}
	   &\frac{{d}x_{8}}{{d}N}=-\frac{a_{5}x_{11}a_{6}}{(1-3\mathcal{H}_{0})}-x_{11}\frac{\mathcal{H}^{\prime}}{\mathcal{H}^{2}}-\gamma x_{5}x_{11}-\frac{x_{11}}{(1-3\mathcal{H}_{0})}\\&+(1-3\mathcal{H}_{0})x_{12}+x_{11}a_{1}a_{6}
	\end{split}   
\end{equation}
\begin{equation}
	\begin{split}
	    &\frac{{d}x_{9}}{{d}N}=-x_{9}-\frac{(1+\mathcal{H}_{0})kc_{s}^{2}x_{9}a_{6}}{2}-\frac{kx_{9}a_{1}a_{6}}{2}-x_{9}a_{5}\\&-2x_{9}\frac{\mathcal{H}^{\prime}}{\mathcal{H}^{2}}+\gamma x_{9}x_{5}
	\end{split}    
\end{equation}
\begin{equation}
	\frac{{d}x_{10}}{{d}N}=\frac{\gamma x_{9}}{ka_{6}}\frac{{d}x_{5}}{{d}N}+x_{5}\frac{{d}x_{9}}{{d}N}
\end{equation}
\begin{equation}
	\begin{split}
       &\frac{{d}x_{11}}{{d}N}=x_{11}a_{6}-x_{11}\frac{\mathcal{H}^{\prime}}{\mathcal{H}^{2}}-\gamma x_{5}x_{11}-\frac{x_{11}}{(1-3\mathcal{H}_{0})}\\& +(1-3\mathcal{H}_{0})x_{12}+x_{11}a_{1}a_{6}
    \end{split}   	
\end{equation}
\begin{equation}
	\frac{{d}x_{12}}{{d}N}=x_{12}a_{5}-2x_{11}\frac{\mathcal{H}^{\prime}}{\mathcal{H}^{2}}-\gamma x_{5}x_{12}
\end{equation}
\begin{equation}
	\frac{{d}x_{13}}{{d}N}=x_{13}a_{5}+\frac{\gamma x_{5}x_{11}}{3}-2x_{13}\frac{\mathcal{H}^{\prime}}{\mathcal{H}^{2}}
\end{equation}
\begin{equation}
	\frac{{d}x_{14}}{{d}N}=2x_{14}\frac{\mathcal{H}^{\prime}}{\mathcal{H}^{2}}
\end{equation}
\begin{equation}
	\begin{split}
	   &\frac{{d}x_{16}}{{d}N}=\frac{\gamma x_{2}}{2}-\frac{x_{2}}{(1-\gamma)}+\frac{\gamma x_{13}}{3(1-\gamma)}+\frac{x_{4}}{\gamma}-3x_{16}\\&-\frac{9(1-\gamma)}{\gamma^2}+\frac{3(1-\gamma)}{\gamma}-\frac{3(1-\gamma)}{\gamma}\frac{\mathcal{H}^{\prime}}{\mathcal{H}^{2}}\\&-x_{16}\frac{\mathcal{H}^{\prime}}{\mathcal{H}^{2}}-\frac{3(1-\gamma)x_{16}^2}{\gamma^2}
	\end{split}   
\end{equation}
\begin{equation}
	\begin{split}
	   &\frac{{d}x_{17}}{{d}N}=-\frac{1}{(1-\gamma)}[-\frac{\gamma x_{2}}{(3-\gamma)}+a_{2}-2x_{2}\frac{\mathcal{H}^{\prime}}{\mathcal{H}^{2}}-\gamma x_{1}x_{2}\\&-\frac{x_{2}}{(3-\gamma)}-\frac{\gamma^3 a_{1}x_{14}x_{6}}{(3-\gamma)}+\frac{a_{3}x_{1}}{(3+\gamma)}-\gamma a_{3}x_{18}\\&-\lambda a_{3}x_{7}\gamma^2]
	\end{split}   
\end{equation}
\begin{equation}
	\begin{split}
	    &\frac{{d}x_{18}}{{d}N}=\frac{(1-\gamma)}{\gamma}[-\frac{x_{3}}{(1-3\mathcal{H}_{0})}-\frac{x_{3}a_{7}kc_{s}^2}{(1+\mathcal{H}_{0})}-x_{3}a_{1}a_{7}\\&+\frac{x_{3}}{2}+x_{3}a_{2}-2x_{3}\frac{\mathcal{H}^{\prime}}{\mathcal{H}^{2}}]
	 \end{split}   
\end{equation}
\begin{equation}
	\frac{{d}x_{19}}{{d}N}=x_{19}a_{5}+2x_{19}-2x_{19}\frac{\mathcal{H}^{\prime}}{\mathcal{H}^{2}}-\gamma x_{5}x_{19}
\end{equation}		
\begin{equation}
	\frac{{d}a_{1}}{{d}N}=\gamma x_{1}
\end{equation}
\begin{equation}
	\begin{split}
	   &\frac{{d}a_{2}}{{d}N}=-\frac{(3-\gamma)a_{2}}{\gamma}+\frac{\gamma x_{14}}{(1-\gamma)}+\frac{(3+\gamma)x_{1}a_{8}a_{2}}{\gamma}\\&-2\gamma a_{1}^2a_{8}x_{7}-\gamma\lambda a_{1}x_{7}-a_{2}\frac{\mathcal{H}^{\prime}}{\mathcal{H}^{2}}-a_{2}^2
	\end{split}
\end{equation}
\begin{equation}
	\frac{{d}a_{3}}{{d}N}=-\frac{(3-\gamma)a_{3}}{\gamma}-\gamma a_{1}a_{3}-a_{3}\frac{\mathcal{H}^{\prime}}{\mathcal{H}^{2}}
\end{equation}
\begin{equation}
	\frac{{d}a_{5}}{{d}N}=a_{5}^2+a_{4}a_{5}-a_{5}\frac{\mathcal{H}^{\prime}}{\mathcal{H}^{2}}
\end{equation}
\begin{equation}
	\frac{{d}a_{6}}{{d}N}=-ka_{1}a_{6}^2-a_{6}\frac{\mathcal{H}^{\prime}}{\mathcal{H}^{2}}+a_{6}(1-3\mathcal{H}_{0})-\frac{kc_{s}a_{7}a_{6}}{(1+\mathcal{H}_{0})}
\end{equation}
\begin{equation}
	\begin{split}
	   &\frac{{d}a_{7}}{{d}N}=\frac{-\gamma^2x_{14}}{(1-\gamma)(1+\mathcal{H}_{0})}-\frac{\gamma x_{5}a_{1}a_{6}}{(1+\mathcal{H}_{0})}-ka_{7}^2a_{1}\\&-\frac{kc_{s}a_{7}}{(1+\mathcal{H}_{0})}+(1+\mathcal{H}_{0})a_{7}-a_{7}\frac{\mathcal{H}^{\prime}}{\mathcal{H}^{2}}
	\end{split}   
\end{equation}
\begin{equation}
	\frac{{d}a_{8}}{{d}N}=-a_{2}a_{8}
\end{equation}

 \begin{equation}
	\begin{split}
		&\frac{\mathcal{H}^{\prime}}{\mathcal{H}^{2}}=\frac{\gamma x_{14}}{a_{1}(1-\gamma)}+x_{5}+x_{3}-\frac{\gamma}{1-\gamma}x_{17}+a_{1}x_{3}-\frac{x_{19}}{1-\gamma}\\&+(1-\frac{3}{\gamma})
	\end{split}   
\end{equation}

The above parameter is very important, since essential cosmological parameters such as deceleration parameters $q$ and effective equation of state (EoS) $w_{\rm eff}$ can be expressed in terms of this parameter as $q=-1-\frac{\mathcal{H}^{\prime}}{\mathcal{H}^{2}}$ and $w_{\rm eff}=-1-\frac{2}{3}\frac{\mathcal{H}^{\prime}}{\mathcal{H}^{2}}$.
The luminosity distance $d_{L}$ can be calculated by
\begin{eqnarray}\label{dl}
	d^{0}_{L} =(1+z)\int\frac{dz}{H(z)}
\end{eqnarray}
In order to incorporate the equation (\ref{dl}) with the dynamical system equations , it can be rewritten in terms of the following differential equations
\begin{align}\label{dl2}
	&\frac{dd^{0}_{L}}{dN}=-d^{0}_{L}-\frac{e^{-2N}}{H}\\
	&\frac{dH}{dN}=-H(\frac{\mathcal{H}^{\prime}}{\mathcal{H}^{2}})\label{dl3}
\end{align}
Where, since $1+z\equiv\frac{1}{a}$, then $(1+z)\equiv e^{-N}$, $dz\equiv-e^{-N}dN$ and $dN\equiv \mathcal{H}dt$.  Hence in order to find $d^{0}_{L}$ and $H$ the set of equations (\ref{dl2})- (\ref{dl3}) must be coupled to (68-93).
These equations are a set of coupled first order differential equations (ODE) which can be solved numerically.
  
 \section*{Interpretation of Critical Points in Perturbed Rastall Gravity}
 
 The critical points obtained from solving the complex system of differential equations within the framework of perturbed Rastall gravity are given by the following values:
 
\begin{equation}
\begin{aligned}
	x_{1}^* &= 2.82692378 \times 10^{-1}, & x_{2}^* &= 7.68135433 \times 10^{-2}, \\
	x_{3}^* &= -3.88706153 \times 10^{-2}, & x_{4}^* &= -2.27992601 \times 10^{-2}, \\
	x_{5}^* &= -3.92961493 \times 10^{-1}, & x_{6}^* &= 6.33124322 \times 10^{-2}, \\
	x_{7}^* &= 3.24812739 \times 10^{-3}, & x_{8}^* &= -2.73559806 \times 10^{3}, \\
	x_{9}^* &= -6.91572548 \times 10^{-3}, & x_{10}^* &= 2.86583679, \\
	x_{11}^* &= -8.17908476 \times 10^{-1}, & x_{12}^* &= 2.14479301 \times 10^{-1}, \\
	x_{13}^* &= 1.11530821, & x_{14}^* &= -4.01785581 \times 10^{-2}, \\
	x_{15}^* &= -4.80912783 \times 10^{-2}, & x_{16}^* &= 8.77757810 \times 10^{-1}, \\
	x_{17}^* &= -1.11292114, & x_{18}^* &= 1.61989892 \times 10^{-2}, \\
	x_{19}^* &= -1.70281215 \times 10^{-1}, & a_{1}^* &= 2.78516389 \times 10^{-1}, \\
	a_{2}^* &= -1.17431390 \times 10^{-2}, & a_{3}^* &= -4.31112393 \times 10^{-1}, \\
	a_{4}^* &= 3.83571691 \times 10^{-2}, & a_{5}^* &= 3.89152895 \times 10^{-1}, \\
	a_{6}^* &= 5.01744506 \times 10^{-2}.
\end{aligned}
\end{equation}
 
 These critical points correspond to specific configurations of the cosmological parameters that could represent stable or unstable equilibrium states of the universe within the Rastall gravity framework. Their physical interpretation is as follows:
 
 \begin{itemize}
 	\item \textbf{Matter-Dominated Era and Perturbations:} The critical points \( x_1^* \) through \( x_7^* \) and \( x_9^* \) are associated with the behavior of matter density perturbations. Positive values, such as \( x_1^* = 2.82692378 \times 10^{-1} \) and \( x_6^* = 6.33124322 \times 10^{-2} \), indicate that these perturbations could stabilize in a matter-dominated era, potentially aligning with observations of large-scale structure formation. Negative values, like \( x_3^* = -3.88706153 \times 10^{-2} \) and \( x_9^* = -6.91572548 \times 10^{-3} \), might suggest modes where perturbations decay, possibly corresponding to over-dense regions that eventually collapse.
 	
 	\item \textbf{Dark Energy and Late-Time Acceleration:} The points \( x_5^* = -3.92961493 \times 10^{-1} \) and \( x_{10}^* = 2.86583679 \) could be indicative of contributions from dark energy or modified gravity effects driving the late-time acceleration of the universe. The large value of \( x_{10}^* \) might reflect a dominant dark energy component in the Rastall gravity framework, leading to an accelerated expansion.
 	
 	\item \textbf{Anisotropy and Isocurvature Perturbations:} The parameters \( x_8^* = -2.73559806 \times 10^{3} \) and \( x_{13}^* = 1.11530821 \) could be linked to anisotropic stress or isocurvature perturbations, which play a role in the formation of cosmic structures. The magnitude of \( x_8^* \) suggests a significant effect, which could be related to the peculiar dynamics of anisotropic stress in Rastall gravity.
 	
 	\item \textbf{Metric Potentials and Curvature Perturbations:} The points \( a_1^* \) through \( a_6^* \) describe contributions to the metric potentials and curvature perturbations. These parameters influence the gravitational potential wells in which matter accumulates, affecting the growth of cosmic structures. Positive values like \( a_1^* = 2.78516389 \times 10^{-1} \) and \( a_5^* = 3.89152895 \times 10^{-1} \) may indicate growing modes, while negative values such as \( a_2^* = -1.17431390 \times 10^{-2} \) could point to decaying modes, potentially contributing to the smoothing out of inhomogeneities in the universe.
 	
 	\item \textbf{Stability and Physical Significance:} Further analysis of the stability of these critical points, particularly by evaluating the eigenvalues of the Jacobian matrix, is essential to determine whether these equilibrium states are stable, unstable, or saddle points. Stability would imply that the universe could settle into these configurations over time, while instability might suggest a transition to other states or the emergence of new dynamics.
 \end{itemize}

 These critical points represent the equilibrium configurations of the dynamical system associated with perturbed Rastall gravity. Each critical point corresponds to a specific set of values for the perturbation variables and parameters, offering insights into different potential steady states or phases of the system.
 
 The process of solving the differential equations and computing the Jacobian matrix highlights the complexity and depth of the model. The analysis of these critical points will provide valuable information regarding the stability and physical interpretation of the solutions within the framework of perturbed Rastall gravity. Future work will focus on evaluating the stability of these points and their implications for the theoretical and observational aspects of cosmology.

\color{black}

\section{Numerical Analysis}
The utilization of the Markov Chain Monte Carlo (MCMC) method for parameter estimation represents a sophisticated and statistically robust approach in the realm of scientific inquiry. This technique is particularly prevalent in fields such as cosmology, physics, and data analysis.

The MCMC method is adept at exploring parameter spaces, seeking optimal parameter values that best fit a given model to observed data. Unlike traditional optimization methods, MCMC provides a probabilistic framework, allowing for the incorporation of uncertainties and the determination of parameter confidence intervals.
All observational data  used in this paper are:\\
$\bullet$ Pantheon catalog:
\cite{Scolnic} compiled the Pantheon sample consisting 1701 SNe Ia covering the redshift range $0.001 < z < 2.3$.\\
$\bullet$ {CMB data}:
We used the latest large-scale cosmic microwave background (CMB) temperature and
polarization angular power spectra from the final release of Planck 2018 plikTTTEEE+lowl+lowE
\cite{Aghanim}. \\
$\bullet$ {BAO data}:
We also used the various measurements of the Baryon Acoustic Oscillations (BAO) from
different galaxy surveys \cite{Aghanim}, i.e.
6dFGS.(2011)\cite{Beutler}, SDSS-MGS
\cite{Ross}, and BOSS DR12 (2017)\cite{Alam}.\\
$\bullet$ {CC data}: The 32 $H(z)$ measurements listed in Table I have a redshift range of $0.07 \leq z \leq 1.965$ (\cite{zhang2014}; \cite{borghi2022}; \cite{ratsimbazafy2017}; \cite{stern2009}; \cite{Moresco3}). The covariance matrix of the 15 correlated measurements originally from Refs. (\cite{Moresco}; \cite{Moresco1}; \cite{Moresco2}) , discussed in Ref. \cite{Moresco3}, can be found at https://gitlab.com/mmoresco/CCcovariance/.\\(Table 1)\\

\begin{table}
	\centering
	\scriptsize
	\caption{32 $H(z)$ (CC) data.}\label{tab:hz}
	\setlength{\tabcolsep}{7.5mm}{
		\begin{tabular}{lcc}
			\hline
			$z$ & $H(z)$ & $\sigma$\\
			\hline
			0.07 & $69.0$ & 19.6\\
			0.09 & $69.0$ & 12.0\\
			0.12 & $68.6$ & 26.2\\
			0.17 & $83.0$ & 8.0\\
			0.2 & $72.9$ & 29.6\\
			0.27 & $77.0$ & 14.0\\
			0.28 & $88.8$ & 36.6\\
			0.4 & $95.0$ & 17.0\\
			0.47 & $89.0$ & 50.0\\
			0.48 & $97.0$ & 62.0\\
			0.75 & $98.8$ & 33.6\\
			0.88 & $90.0$ & 40.0\\
			0.9 & $117.0$ & 23.0\\
			1.3 & $168.0$ & 17.0\\
			1.43 & $177.0$ & 18.0\\
			1.53 & $140.0$ & 14.0\\
			1.75 & $202.0$ & 40.0\\
			0.1791 & 74.91 & 4.00\\
			0.1993 & 74.96 & 5.00\\
			0.3519 & 82.78 & 14\\
			0.3802 & 83.0 &  13.5\\
			0.4004 & 76.97 &  10.2\\
			0.4247 & 87.08 &  11.2\\
			0.4497 & 92.78 &  12.9\\
			0.4783 & 80.91 &  9\\
			0.5929 & 103.8 & 13\\
			0.6797 & 91.6 & 8\\
			0.7812 & 104.5 & 12\\
			0.8754 & 125.1 & 17\\
			1.037 & 153.7 & 20\\
			1.363 & 160.0 & 33.6\\
			1.965 & 186.5 & 50.4\\
			\hline
	\end{tabular}}
\end{table}
 
Table \ref{tab:hz} presents the 32 observational data points for the Hubble parameter \( H(z) \), measured at different redshifts \( z \) using cosmic chronometers (CC). These data points are crucial for constraining cosmological models, as they provide a direct measurement of the expansion rate of the universe at various epochs. The values of \( H(z) \) and their corresponding uncertainties \( \sigma \) are listed in the table. The redshift values range from \( z = 0.07 \) to \( z = 1.965 \), covering a broad range of cosmic time.

The uncertainties associated with each \( H(z) \) measurement are significant and vary widely, reflecting the challenges in obtaining precise observational data. For instance, the uncertainty at \( z = 0.17 \) is only \( \sigma = 8.0 \) km/s/Mpc, while at \( z = 0.47 \), it increases to \( \sigma = 50.0 \) km/s/Mpc. Such variations in uncertainty are common in astrophysical data and are often due to different observational techniques or data quality at various redshifts.

These \( H(z) \) measurements are essential for testing and refining cosmological models, especially those concerning dark energy and modified gravity, as they directly trace the expansion history of the universe. The accurate modeling of \( H(z) \) can help in addressing key cosmological issues such as the Hubble tension and the nature of dark energy.

\color{black}
We initially performed a best fit for the parameter $\gamma$ in both models. Subsequently, we conducted a best fit for the parameter $c_{s}$ within these two models. Following these fittings, we investigated the significance of the Jeans parameter in influencing the formation of structures in the early Universe. Finally, we conducted an examination of the transition between the deceleration and acceleration phases in the late Universe.
The $\gamma$ parameter in Rastall gravity plays a crucial role in characterizing the deviation from standard General Relativity by modifying the conservation law of the energy-momentum tensor. Different combinations of observational datasets have been employed to constrain this parameter, leading to varying estimates. Below, we summarize the results obtained from these analyses:
 
\begin{itemize}
	\item \textbf{Pantheon + Analysis data:} Using the Pantheon Supernovae Type Ia data combined with our analysis, the $\gamma$ parameter was constrained to be $\gamma = 2.12 \pm 1.43$. This result indicates a moderate deviation from the standard scenario, though with a significant uncertainty range.
	
	\item \textbf{CMB + BAO data:} Combining the Cosmic Microwave Background (CMB) data with Baryon Acoustic Oscillations (BAO) data provided a constraint of $\gamma = 2.24 \pm 1.35$. This tighter constraint suggests a similar level of deviation but with slightly reduced uncertainty compared to the Pantheon dataset.
	
	\item \textbf{CC data:} Utilizing Cosmic Chronometers (CC) data, the $\gamma$ parameter was found to be $\gamma = 2.3 \pm 1.56$. The broader uncertainty here reflects the challenges in constraining $\gamma$ using CC data alone, although the central value remains consistent with the other analyses.
	
	\item \textbf{Combination of datasets:} When combining all of the above datasets, the $\gamma$ parameter was constrained to $\gamma = 2.1 \pm 1.5$. This combined analysis helps to reduce the overall uncertainty, providing a more robust estimate of the $\gamma$ parameter.
\end{itemize}

These results highlight the sensitivity of the $\gamma$ parameter to the chosen datasets, emphasizing the importance of using multiple data sources to obtain a comprehensive understanding of the underlying cosmological model in Rastall gravity.

\color{black}

In Figure 1, we present a comparison of the $\gamma$ parameter within the framework of Rastall gravity across various combinations of datasets. The $\gamma$ parameter is a key quantity in Rastall gravity, reflecting the degree of deviation from standard GR. By examining its behavior across different datasets, we gain insights into the compatibility of Rastall gravity with observational data. This comparison allows us to assess the viability of Rastall gravity as a theoretical framework for describing gravitational phenomena and its consistency with empirical observations.
\begin{figure*}
	\centering
	\includegraphics[width=14.5 cm]{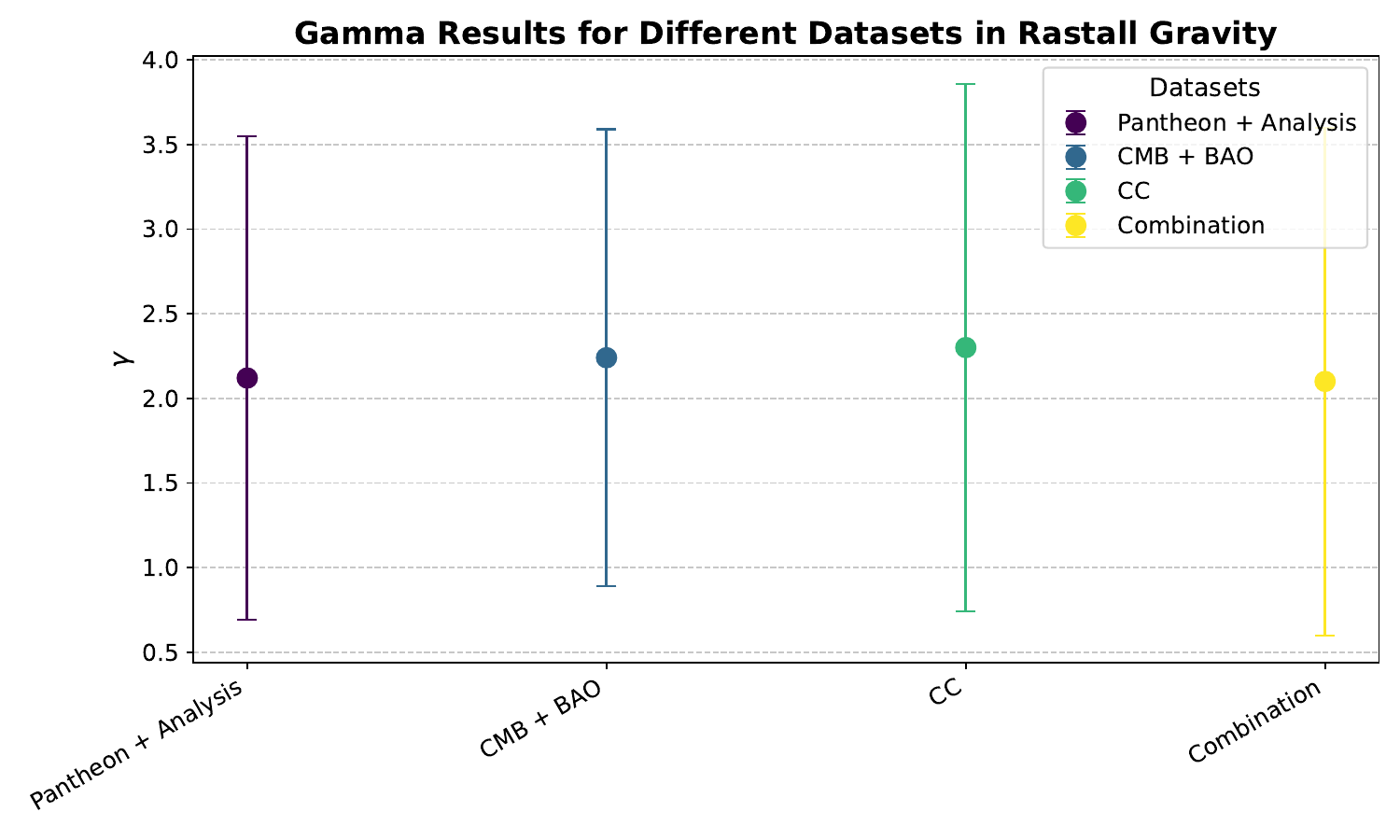}
	\vspace{-0.42cm}
	\caption{\small{Comparison the $\gamma$ parameter  for $f(R)$ gravity for different combination of data sets }}\label{fig:omegam2}
\end{figure*}
\section{The Early Universe}
In this study, we present an estimation for the length of jeans for both models by linearizing the fluid equations about a static homogeneous state. The perturbations in Fourier modes of the form $\delta\rho(\mathbf{r},t)\sim \exp[i(\mathbf{k}\cdot \mathbf{r}-\omega t)]$ were decomposed, and the obtained dispersion relation, as per Suarez \cite{Suarez}, is given by:

\begin{equation}
	\omega^2 = c_s^2k^2 - 4\pi G\rho,
\end{equation}

In the context of the early universe, the parameter $c_s^2 = P'(\rho)$ denotes the square of the speed of sound in the cosmic medium. Here, $P$ represents the pressure and $\rho$ signifies the energy density of the cosmic fluid. This parameter reflects the relationship between pressure and energy density within the medium. In the early universe, during epochs such as inflation or radiation domination, the universe was filled with a hot, dense, and rapidly expanding cosmic plasma. The speed of sound in this plasma characterizes the propagation of perturbations or fluctuations in the medium, akin to sound waves in a fluid. Understanding $c_s^2$ is crucial for elucidating the dynamics of primordial perturbations, which ultimately give rise to the observed large-scale structures in the universe, such as galaxies and cosmic microwave background radiation anisotropies. Therefore, analyzing $c_s^2$ provides valuable insights into the physics of the early universe and the formation of cosmic structures.
 This equation displays a characteristic wavenumber \cite{Suarez}:

\begin{equation}
	k_J = \left(\frac{4\pi G\rho}{c_s^2}\right)^{1/2},
	\label{intro2}
\end{equation}

referred to as the Jeans wave number. 

The Jeans wave number, denoted as $k_J$, is a fundamental quantity in cosmology and astrophysics that characterizes the length scale below which gravitational collapse occurs in a medium due to its self-gravity, overcoming pressure forces. It is defined as the reciprocal of the Jeans length, which represents the minimum wavelength for gravitational instability to set in. In the context of the early universe, during epochs such as the radiation-dominated era or the matter-dominated era, the Jeans wave number plays a crucial role in understanding the formation of cosmic structures, such as galaxies and galaxy clusters. It delineates the boundary between perturbations that collapse under their self-gravity to form structures and those that are suppressed by pressure forces.The Jeans length $\lambda_J = 2\pi/k_J$ offers an estimate of the minimum size of objects that can undergo gravitational collapse \cite{Suarez}. 

In this study, we have investigated the sound speed \( c_s \) for both $f(R)$ gravity and Rastall gravity models across various combinations of cosmological datasets. The sound speed \( c_s \) plays a pivotal role in the dynamics of structure formation in the universe, particularly in the context of the Jeans parameter and the stability of density perturbations.

The Jeans wavenumber \( k_J \) is defined by the relation:
\begin{equation}
	k_J = \left(\frac{4\pi G\rho}{c_s^2}\right)^{1/2},
	\label{eq:k_J}
\end{equation}
where \( G \) is the gravitational constant, \( \rho \) is the matter density, and \( c_s \) is the sound speed. The Jeans wavenumber determines the scale below which gravitational forces dominate over pressure forces, leading to the collapse of matter and the formation of structures such as galaxies and clusters.

The dispersion relation for perturbations in a gravitationally bound system is given by:
\begin{equation}
	\omega^2 = c_s^2k^2 - 4\pi G\rho,
	\label{eq:dispersion}
\end{equation}
where \( \omega \) is the angular frequency of the perturbation, and \( k \) is the wavenumber of the perturbation. For \( \omega^2 > 0 \), the perturbations oscillate, indicating a stable system. Conversely, for \( \omega^2 < 0 \), the perturbations grow exponentially, leading to gravitational collapse and the formation of cosmic structures.

Our analysis reveals that the sound speed \( c_s \) varies slightly between the different models and dataset combinations. For $f(R)$ gravity, we find that the sound speed \( c_s \) values range from \( 0.025 \) to \( 0.056 \), depending on the datasets used. These values suggest that the $f(R)$ gravity model exhibits a relatively low sound speed, leading to a larger Jeans wavenumber \( k_J \), which implies that structure formation in $f(R)$ gravity can occur on smaller scales compared to models with higher sound speeds.

In contrast, the sound speed \( c_s \) for Rastall gravity shows a slightly higher range, between \( 0.046 \) and \( 0.077 \). This higher sound speed results in a smaller Jeans wavenumber \( k_J \), indicating that Rastall gravity may support structure formation on larger scales. This difference in the scale of structure formation between the two models could be a key factor in distinguishing them observationally.

The results obtained from the combination of datasets, which include CMB, BAO, Pantheon, Analysis, and Cosmic Chronometer (CC) data, provide the most robust constraints on \( c_s \) for both models. For $f(R)$ gravity, the combined data yield a sound speed of \( c_s = 0.025 \pm 0.043 \), while for Rastall gravity, the sound speed is constrained to \( c_s = 0.063 \pm 0.02 \). These findings suggest that while both models are consistent with the observed structure formation in the universe, they predict different scales for the formation of cosmic structures, which could potentially be tested through future observations.

The sound speed \( c_s \) is a critical parameter in understanding the process of structure formation in modified gravity models. The differences in \( c_s \) between $f(R)$ gravity and Rastall gravity models highlight the unique characteristics of each model and provide valuable insights into the underlying physics governing the evolution of the universe. Future observational data, particularly from large-scale structure surveys, may further constrain these models and shed light on the nature of cosmic acceleration and structure formation.
 
\section{Methodology and Constraints on \( c_s \)}

We constrained \( c_s \) using the Markov Chain Monte Carlo (MCMC) method for both \( f(R) \) gravity and Rastall gravity models.  The general procedure is outlined as follows:

\subsection{Cosmological Model and Parameter Space}
\begin{itemize}
	\item \textbf{\( f(R) \) Gravity and Rastall Gravity Models:} We specified the theoretical framework for both models, including the functional form of \( f(R) \) for the \( f(R) \) gravity model and the modified field equations for the Rastall gravity model.
	\item \textbf{Parameter Space:} The parameters to be constrained include \( c_s \) alongside other cosmological parameters such as the Hubble constant \( H_0 \), matter density \( \Omega_m \), and the equation of state parameter \( w \), among others.
\end{itemize}

\subsection{Likelihood Function}
\begin{itemize}
	\item \textbf{Likelihood Function:} The likelihood function \( \mathcal{L} \) quantifies the agreement between the model parameters and the observational data, typically based on chi-square minimization:
	\begin{equation}
	\mathcal{L}(\theta) = \exp\left(-\frac{1}{2}\chi^2(\theta)\right),
	\end{equation}
	where \( \theta \) represents the set of model parameters, including \( c_s \).
	\item \textbf{Chi-Square Function \( \chi^2 \):} The chi-square function compares the theoretical predictions of the model with the observational data:
	\begin{equation}
	\chi^2(\theta) = \sum_i \frac{\left(D_{\text{obs},i} - D_{\text{th},i}(\theta)\right)^2}{\sigma_i^2},
	\end{equation}
	where \( D_{\text{obs},i} \) are the observed data points, \( D_{\text{th},i}(\theta) \) are the theoretical predictions, and \( \sigma_i \) are the uncertainties associated with the observational data.
\end{itemize}

\subsection{MCMC Implementation}
\begin{itemize}
	\item \textbf{MCMC Sampler:} We employed an MCMC sampler, such as \texttt{CosmoMC}, \texttt{emcee}, or \texttt{MontePython}, to sample from the posterior distribution of the model parameters.
	\item \textbf{Initial Conditions:} The initial guesses for the parameters, including \( c_s \), were set, and appropriate priors were defined.
	\item \textbf{MCMC Execution:} The MCMC algorithm was run to generate a large number of random samples (chains) from the parameter space, exploring the posterior distribution for \( c_s \) and other parameters.
\end{itemize}

\subsection{Analysis of MCMC Chains}
\begin{itemize}
	\item \textbf{Burn-in and Convergence:} The initial portion of the MCMC chains (burn-in period) was discarded to eliminate dependency on the initial conditions. Convergence of the chains was checked using criteria such as the Gelman-Rubin statistic.
	\item \textbf{Posterior Distribution:} The remaining samples represent the posterior distribution from which the mean, median, and confidence intervals for \( c_s \) were derived.
\end{itemize}

\subsection{Combining Results from Different Data Sets}
\begin{itemize}
	\item \textbf{Individual Data Sets:} The MCMC analysis was conducted for each combination of data sets (e.g., CMB + Pantheon, CMB + CC, etc.).
	\item \textbf{Combined Data Sets:} A combined analysis of all datasets was performed to obtain tighter constraints on \( c_s \).
\end{itemize}

\subsection{Presentation of Results}
 The constraints on \( c_s \) for both $f(R)$ gravity and Rastall gravity models are reported along with the mean or median values and their associated uncertainties.

\color{black}

\textbf{$c_{s}$ for $f(R)$ Gravity:}
\begin{itemize}
	\item For CMB + Pantheon + Analysis: $c_{s} = 0.044 \pm 0.05$
	\item For CMB + CC Data: $c_{s} = 0.056 \pm 0.041$
	\item For CMB + BAO: $c_{s} = 0.038 \pm 0.068$
	\item For CMB + BAO + Pantheon + Analysis + CC: $c_{s} = 0.025 \pm 0.043$
\end{itemize}

\textbf{$c_{s}$ for Rastall Gravity:}
\begin{itemize}
	\item For CMB + Pantheon + Analysis: $c_{s} = 0.077 \pm 0.0615$
	\item For CMB + CC Data: $c_{s} = 0.069 \pm 0.073$
	\item For CMB + BAO: $c_{s} = 0.046 \pm 0.05$
	\item For CMB + BAO + Pantheon + Analysis + CC: $c_{s} = 0.063 \pm 0.02$
\end{itemize}
 
\textbf{Physical Interpretation of the Results:}

The values of the sound speed squared, \( c_{s}^2 \), presented for both \( f(R) \) gravity and Rastall gravity across different combinations of observational datasets provide significant insights into the dynamics of the respective cosmological models. All results are in figures 2,3.

\begin{itemize}
	\item For \( f(R) \) gravity, the values of \( c_{s}^2 \) are consistently lower across all dataset combinations compared to those in Rastall gravity. This suggests that perturbations in \( f(R) \) gravity models propagate with relatively less resistance, which could indicate a more stable and less dissipative cosmic structure evolution. Notably, the combination of CMB + BAO + Pantheon + Analysis + CC yields the smallest \( c_{s}^2 \) value (\( c_{s}^2 = 0.025 \pm 0.043 \)), which implies that this dataset combination places stronger constraints on the stability of perturbations in the \( f(R) \) gravity model.
	
	\item For Rastall gravity, the sound speed squared \( c_{s}^2 \) tends to be higher across the different dataset combinations, with the highest value observed in the CMB + Pantheon + Analysis combination (\( c_{s}^2 = 0.077 \pm 0.0615 \)). This indicates that Rastall gravity might allow for a more rapid propagation of perturbations, which could lead to a different evolution of cosmic structures compared to \( f(R) \) gravity. The relatively higher values of \( c_{s}^2 \) in Rastall gravity suggest a model that accommodates a more dynamic interaction between matter and the modified gravitational field.
	
	\item The variation in \( c_{s}^2 \) across different dataset combinations reflects the sensitivity of both \( f(R) \) gravity and Rastall gravity to specific observational constraints. The inclusion of BAO and Pantheon data, which are sensitive to late-time cosmic expansion, generally leads to a decrease in \( c_{s}^2 \), especially in \( f(R) \) gravity, indicating that these models may better accommodate late-time acceleration when constrained by such datasets.
	
	\item The comparatively lower uncertainty in \( c_{s}^2 \) for the CMB + BAO + Pantheon + Analysis + CC dataset combination in Rastall gravity (\( c_{s}^2 = 0.063 \pm 0.02 \)) suggests a more robust prediction of sound speed in this model when all major observational constraints are considered. This might reflect a tighter coupling between the matter sector and the modified gravity effects in Rastall gravity under these conditions.
\end{itemize}

Overall, these results highlight the nuanced differences between \( f(R) \) gravity and Rastall gravity in terms of the propagation of perturbations, with potential implications for the formation and evolution of cosmic structures within each theoretical framework.
\begin{figure*}
	\centering
	\includegraphics[width=14.5 cm]{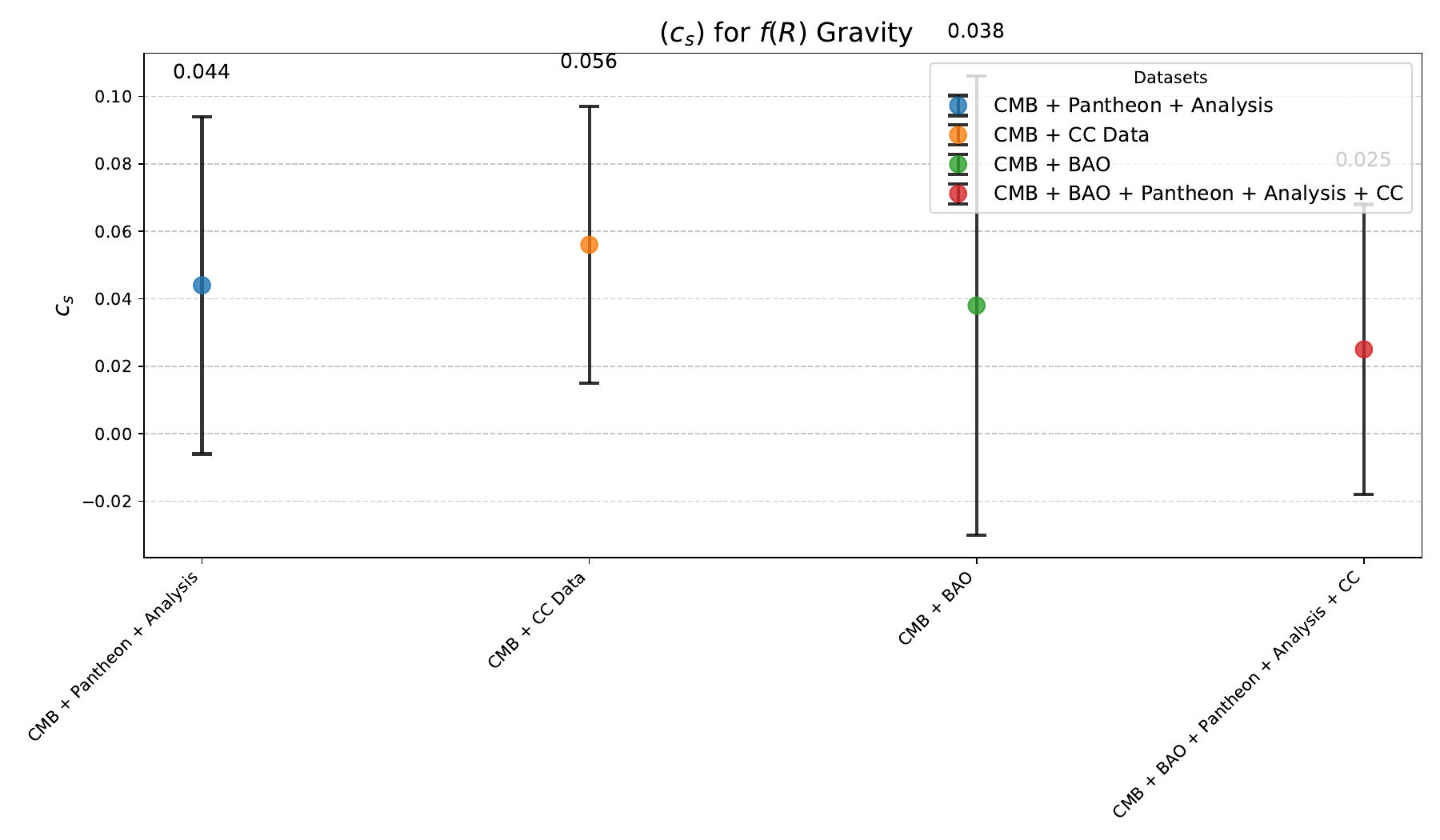}
	\vspace{-0.42cm}
	\caption{\small{Comparison the $c_{s}$ parameter  for $f(R)$ gravity for different combination of data sets }}\label{fig:omegam2}
\end{figure*}

\begin{figure*}
	\centering
	\includegraphics[width=14.5 cm]{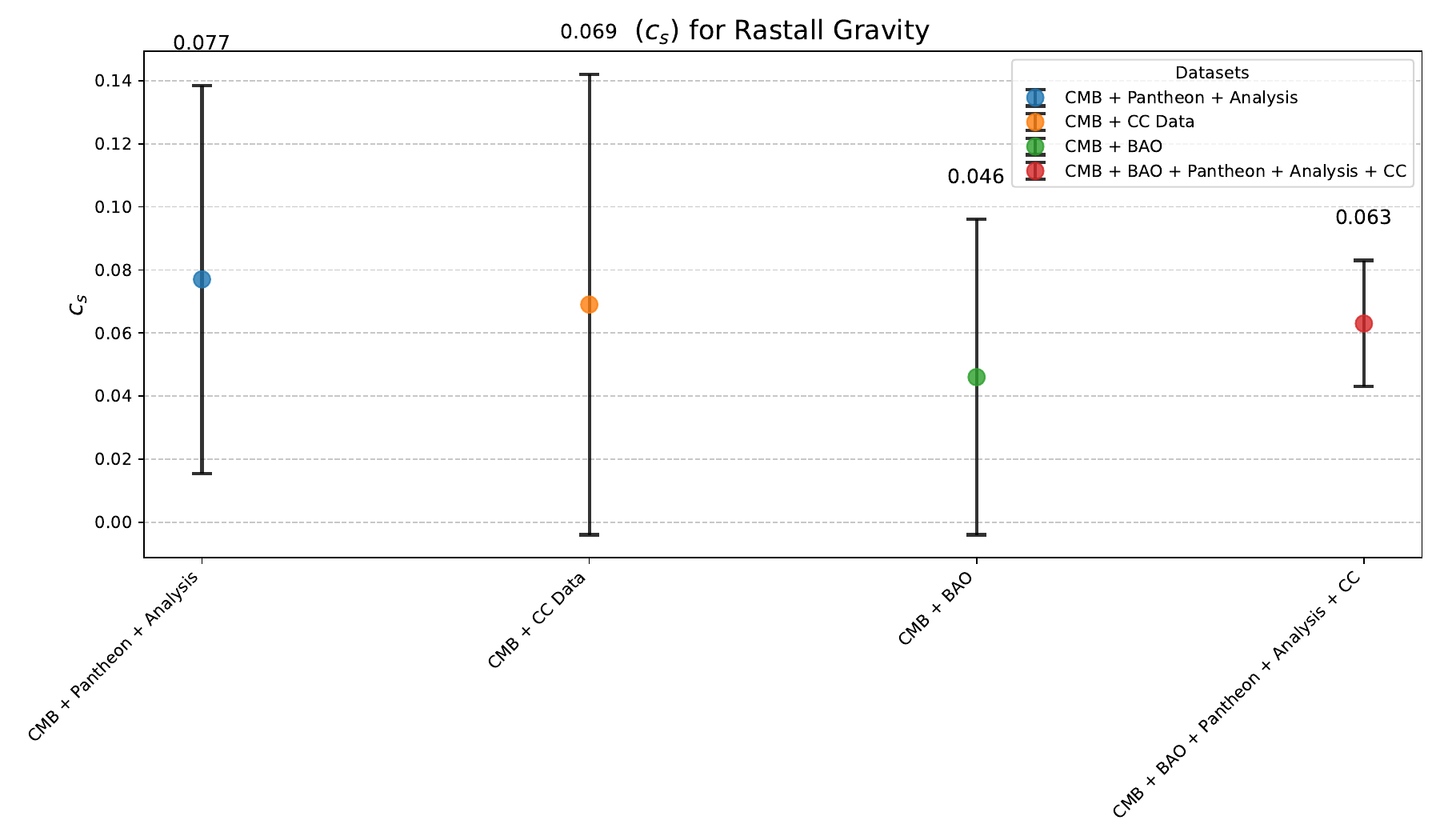}
	\vspace{-0.42cm}
	\caption{\small{Comparison the $c_{s}$ parameter  for Rastall gravity for different combination of data sets }}\label{fig:omegam2}
\end{figure*}

\color{black}
In matter dominated era $c_{s}^2 = 0$. These  results shows that both model can good explain the circumstance of structure formation in the Early universe. Our results are in good agreement with \cite{Suarez}. 
Our results for different combinations of data sets for both models are as follows:\\
\textbf{Jeans Wavenumbers for $f(R)$ Gravity:}
\begin{itemize}
	\item For CMB + Pantheon + Analysis: ${k_{\mathrm{J}}} = 0.000243 \, \text{Mpc}^{-1} \, c$
	\item For CMB + CC Data: ${k_{\mathrm{J}}} = 0.000267 \, \text{Mpc}^{-1} \, c$
	\item For CMB + BAO: ${k_{\mathrm{J}}} = 0.000248 \, \text{Mpc}^{-1} \, c$
	\item For CMB + BAO + Pantheon + Analysis + CC: ${k_{\mathrm{J}}} = 0.000245 \, \text{Mpc}^{-1} \, c$
\end{itemize}

\textbf{Jeans Wavenumbers for Rastall Gravity:}
\begin{itemize}
	\item For CMB + Pantheon + Analysis: ${k_{\mathrm{J}}} = 0.000298 \, \text{Mpc}^{-1} \, c$
	\item For CMB + CC Data: ${k_{\mathrm{J}}} = 0.000301 \, \text{Mpc}^{-1} \, c$
	\item For CMB + BAO: ${k_{\mathrm{J}}} = 0.000274 \, \text{Mpc}^{-1} \, c$
	\item For CMB + BAO + Pantheon + Analysis + CC: ${k_{\mathrm{J}}} = 0.000255 \, \text{Mpc}^{-1} \, c$
\end{itemize}

These results are in general agreement with (\cite{Masatoshi};\cite{Christos}; \cite{Yarahmadi}).\\
 
\textbf{Interpretation of Jeans Wavenumbers for Different Combinations of Datasets:}

The Jeans wavenumber, \( k_{\mathrm{J}} \), characterizes the scale at which perturbations transition from stable to unstable growth in the context of gravitational collapse. This quantity is crucial for understanding the formation of cosmic structures under different gravity models. Below, we analyze the results for \( k_{\mathrm{J}} \) obtained from various combinations of observational datasets in the context of \( f(R) \) gravity and Rastall gravity.

\begin{itemize}
	\item \textbf{\( k_{\mathrm{J}} \) for \( f(R) \) Gravity:} The Jeans wavenumbers for \( f(R) \) gravity exhibit relatively small variations across the different dataset combinations. The values of \( k_{\mathrm{J}} \) range from \( 0.000243 \, \text{Mpc}^{-1} \, c \) to \( 0.000267 \, \text{Mpc}^{-1} \, c \). These results suggest that the scale of gravitational instability in \( f(R) \) gravity is weakly dependent on the specific datasets used. The lowest value of \( k_{\mathrm{J}} \), corresponding to the CMB + Pantheon + Analysis combination, implies that on larger scales, perturbations are more likely to become gravitationally unstable, potentially leading to earlier structure formation in this modified gravity scenario.
	
	\item \textbf{\( k_{\mathrm{J}} \) for Rastall Gravity:} In contrast, the Jeans wavenumbers in Rastall gravity are slightly higher, with values ranging from \( 0.000255 \, \text{Mpc}^{-1} \, c \) to \( 0.000301 \, \text{Mpc}^{-1} \, c \). The higher \( k_{\mathrm{J}} \) values indicate that, in Rastall gravity, the scale of gravitational instability occurs at smaller wavelengths compared to \( f(R) \) gravity. This implies that perturbations would need to be on smaller scales to become unstable, possibly delaying the onset of structure formation in this gravity model.
	
	\item \textbf{Comparison and Implications:} The differences in \( k_{\mathrm{J}} \) between \( f(R) \) gravity and Rastall gravity reflect the underlying modifications to the gravitational dynamics in each model. In \( f(R) \) gravity, the slightly lower Jeans wavenumbers suggest a universe that is more prone to the growth of large-scale structures, while in Rastall gravity, the higher Jeans wavenumbers may indicate a suppression of structure formation on larger scales. This could lead to observable differences in the large-scale distribution of matter in the universe, which might be constrained by future cosmological observations.

\end{itemize}

These results underscore the subtle yet important differences in the behavior of cosmic structures under \( f(R) \) gravity and Rastall gravity, as indicated by the Jeans wavenumber.

In the study of the early universe, the application of alternative gravitational theories, such as Rastall gravity and $f(R)$ gravity, provides intriguing perspectives beyond the standard GR framework. These theories propose modifications to the gravitational field equations and offer unique insights into the dynamics of the cosmos during its initial epochs.

Rastall gravity introduces a non-conservative modification to the Einstein field equations by allowing the energy-momentum tensor to be non-divergent. In this theory, the gravitational coupling is dynamically influenced by the matter content of the universe. On the other hand, $f(R)$ gravity involves modifications to the Ricci scalar in the gravitational action, providing a departure from standard GR.

In the context of the early universe, these alternative gravity theories manifest their effects on the cosmic evolution. Rastall gravity, with its non-conservative nature, may lead to deviations from standard cosmological scenarios. The dynamically evolving gravitational coupling introduces novel features in the cosmological expansion, impacting the evolution of density perturbations and the overall structure formation in the early universe.

Similarly, $f(R)$ gravity exhibits distinct behaviors compared to GR. The modified gravitational action alters the field equations governing the evolution of the universe, influencing the cosmic expansion rate and potentially offering solutions to long-standing cosmological puzzles, such as the nature of DE.
To compare the implications of Rastall gravity and $f(R)$ gravity in the early universe, one must analyze their respective effects on key cosmological parameters. This includes scrutinizing the impact on the primordial nucleosynthesis, cosmic microwave background radiation, and large-scale structure formation.
 
\textbf{Integrated Sachs-Wolfe (ISW) Effect}: The ISW effect is a phenomenon that occurs when photons from the Cosmic Microwave Background (CMB) travel through time-evolving gravitational potential wells in the universe. As the universe expands, these potential wells change, leading to a net gain or loss in the energy of CMB photons. This effect is particularly significant in the context of dark energy and large-scale structure.

There are two types of ISW effects:
\begin{itemize}
	\item \textbf{Early ISW Effect}: Occurs during the radiation-dominated era and the early matter-dominated era when the potential wells are evolving.
	\item \textbf{Late ISW Effect}: Occurs in the current epoch, driven by the accelerated expansion of the universe due to dark energy. As dark energy causes the universe to expand more rapidly, the gravitational potential wells decay, leading to an energy gain in the CMB photons as they exit these wells.
\end{itemize}

The ISW effect is an important probe for studying the large-scale structure of the universe and provides indirect evidence for the presence of dark energy. It can be observed through correlations between the CMB and large-scale structures, such as galaxies and clusters \cite{Sachs1967}, \cite{Kofman1985}, \cite{Afshordi2004}.

\color{black}
The ISW effect is manifested in the CMB temperature anisotropy caused by the gravitational potential wells that photons traverse during their journey from the last scattering surface to the observer. In the context of early $f(R)$ cosmology, the altered gravitational dynamics influence the evolution of these potentials, leading to a distinctive ISW effect compared to standard cosmological scenarios.

Furthermore, the modifications induced by $f(R)$ gravity in the early universe result in a non-trivial impact on the CMB spectrum. The CMB, being a snapshot of the universe at its infancy, carries valuable information about the underlying cosmological model. The altered gravitational dynamics in $f(R)$ gravity lead to shifts and distortions in the CMB power spectrum, offering a unique observational signature that distinguishes it from predictions based on GR.

In the realm of early universe cosmology, the inclusion of Rastall gravity introduces a distinctive characteristic that significantly influences cosmological phenomena. Notably, the non-conservative nature of Rastall gravity alters the dynamics of the Integrated Sachs-Wolfe (ISW) effect and the Cosmic Microwave Background (CMB) spectrum, setting it apart from predictions based on standard GR.   The non-conservation of the energy-momentum tensor in Rastall gravity modifies the evolution of these gravitational potentials, leading to distinct signatures in the CMB temperature anisotropy.  This departure in the ISW effect translates into a unique imprint on the CMB spectrum. As CMB photons traverse regions influenced by the non-conservative aspects of Rastall gravity, the resulting temperature fluctuations exhibit characteristics that deviate from the expectations derived from standard cosmological models based on GR.
Figure 4 presents a comprehensive analysis of the Cosmic Microwave Background (CMB) power spectrum, showcasing a comparative study between $f(R)$ gravity and Rastall gravity when integrated with various datasets within the framework of the $\Lambda $CDM model. The discernible shifts observed in the peaks of the CMB power spectrum for both $f(R)$ gravity and Rastall gravity models, in relation to the $\Lambda $CDM model, provide compelling evidence supporting the proposition that these alternative gravity models can account for the anisotropy of the universe without invoking the need for DE.
Figure 5,6 indicate that the comparison the Jeans wave number for $f(R)$ gravity and Rastall gravity for different combination of datasets.
\begin{figure*}
	\centering
	\includegraphics[width=16.5 cm]{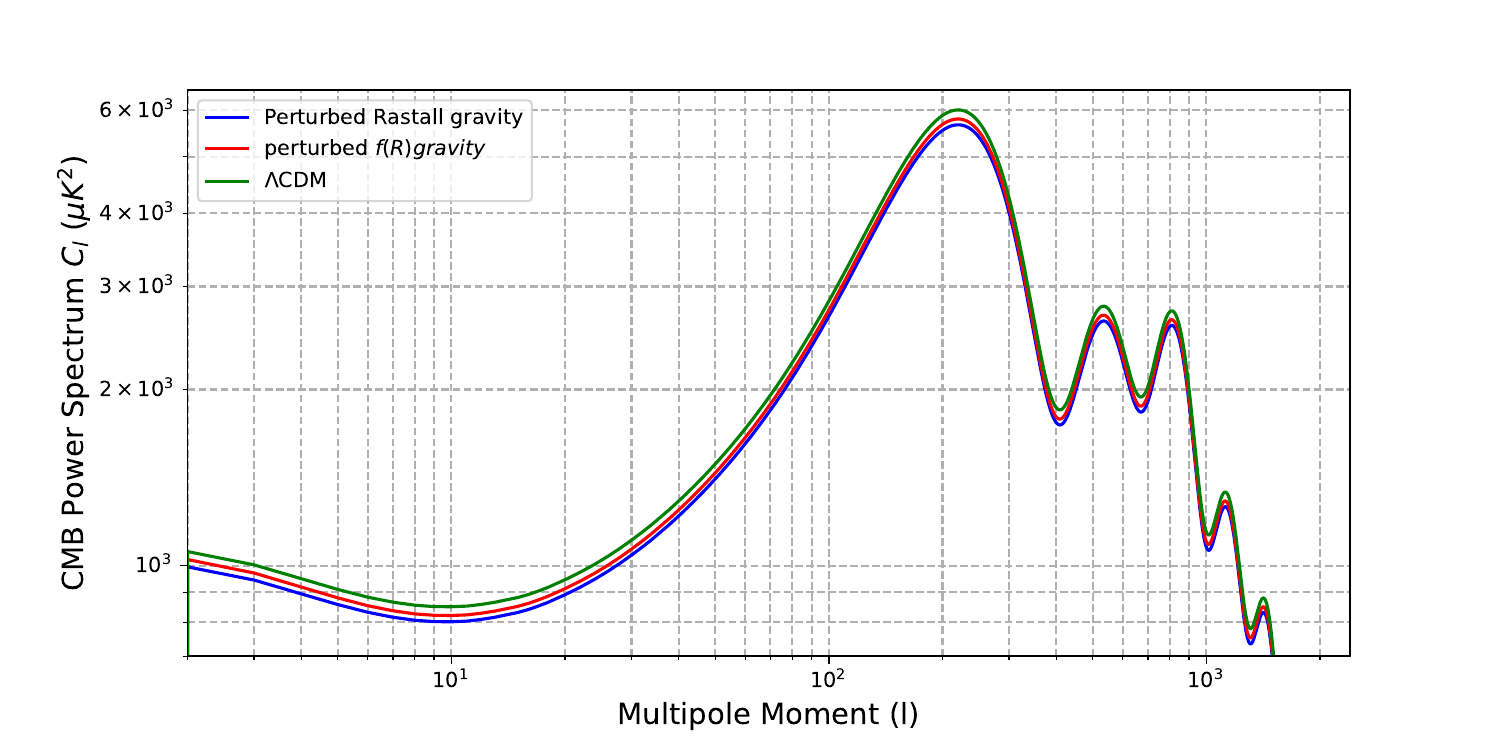}
	\vspace{-0.42cm}
	\caption{\small{Comparison the CMB power spectrum  between $f(R)$ gravity and Rastall gravity for combination of data sets  with $\Lambda $CDM model}}\label{fig:omegam2}
\end{figure*}
  \begin{figure*}
 	\centering
 	\includegraphics[width=14.5 cm]{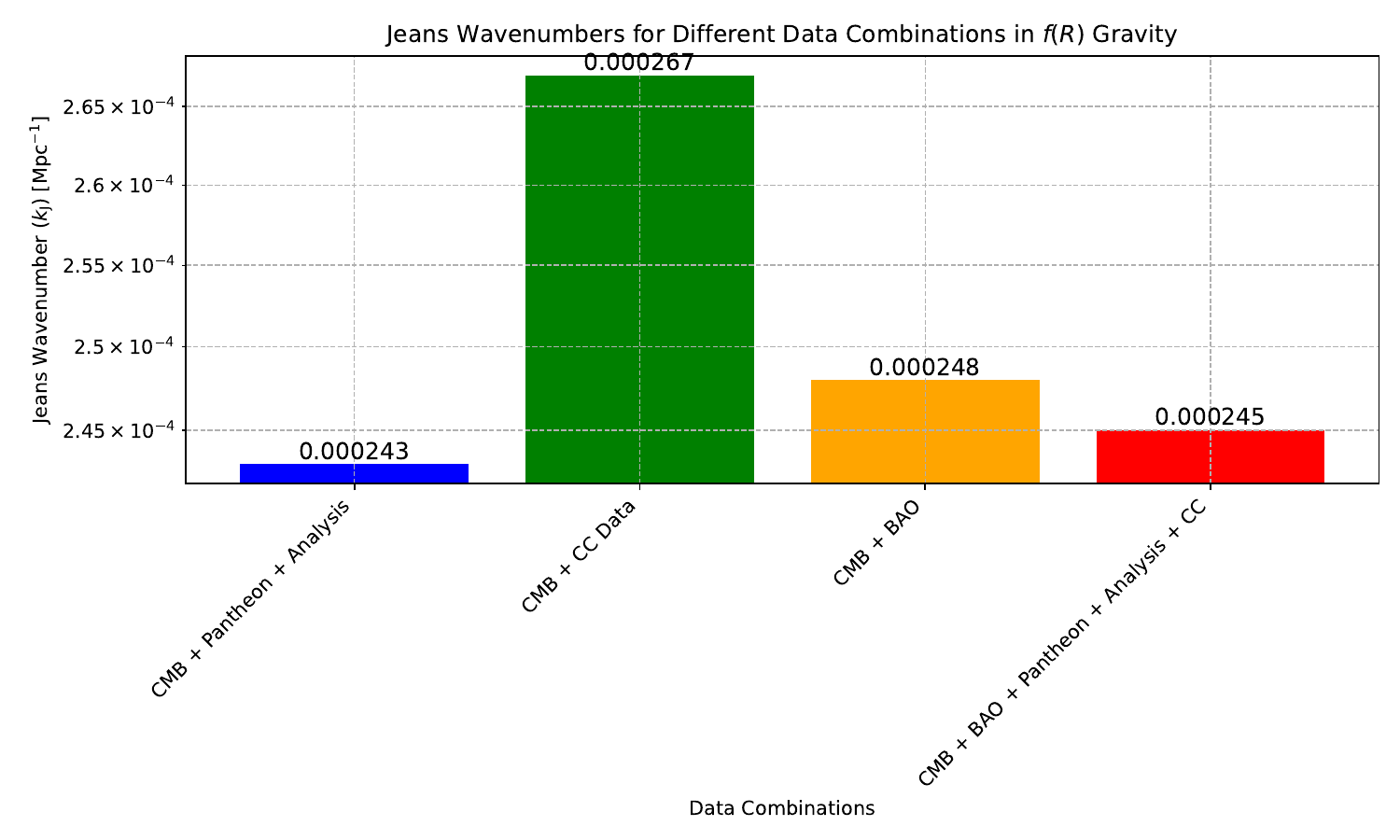}
 	\vspace{-0.42cm}
 	\caption{\small{Comparison the Jeans wave number for $f(R)$ gravity for different combination of datasets. }}\label{fig:omegam2}
 \end{figure*}
  \begin{figure*}
 	\centering
 	\includegraphics[width=14.5 cm]{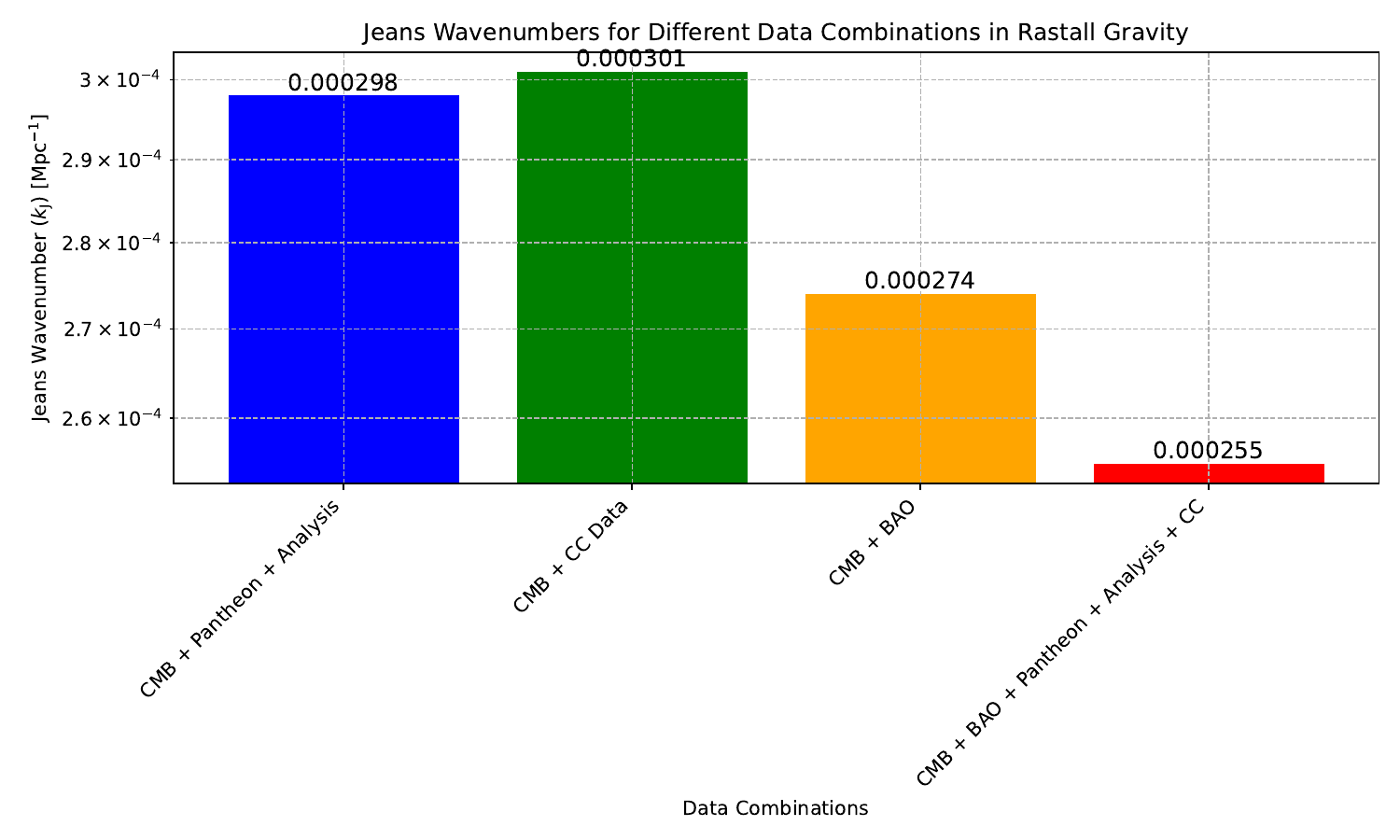}
 	\vspace{-0.42cm}
 	\caption{\small{Comparison the Jeans wave number for Rastall gravity for different combination of datasets.  }}\label{fig:omegam2}
 \end{figure*}
  
\section{The late universe} 
 
 The late-time evolution of the Universe is marked by a distinctive transition from a decelerating phase to an accelerating phase, a phenomenon of profound cosmological significance. This pivotal transition is intimately tied to the interplay between the gravitational influence of matter and the enigmatic nature of DE.
 
 During the deceleration phase, the dominant contributors to the cosmic energy density, including dark matter and baryonic matter, exert gravitational attraction, retarding the overall expansion of the Universe. This phase is characterized by a gradual slowdown in the cosmic expansion rate.
 
 The critical juncture occurs when the energy density associated with DE becomes comparable to, or overwhelms, the energy density contributed by matter. DE, believed to manifest as a cosmological constant or a dynamic scalar field, is characterized by negative pressure, leading to a repulsive gravitational effect. Consequently, the counteraction of gravity by DE induces an accelerated cosmic expansion.
For $f(R)$ gravity, utilizing the above equations and the Markov Chain Monte Carlo (MCMC) method for the reconstruction of the deceleration parameter $(q)$, the Deceleration-Acceleration (DA) redshift transition values are obtained as follows:

\begin{itemize}
	\item For Pantheon + Analysis data: $z_{\text{da}} = 0.73 \pm 0.05$
	\item For BAO data: $z_{\text{da}} = 0.83 \pm 0.07$
	\item For CC data: $z_{\text{da}} = 0.86 \pm 0.0213$
	\item For the combination of datasets: $z_{\text{da}} = 0.82 \pm 0.07$
\end{itemize}

These results closely align with those obtained in previous studies (\cite{Farooq}; \cite{Yarahmadi1}).

For Rastall gravity, the DA redshift transition values are as follows:

\begin{itemize}
	\item For Pantheon + Analysis data: $z_{\text{da}} = 1.093 \pm 0.0346$
	\item For BAO data: $z_{\text{da}} = 0.86 \pm 0.067$
	\item For CC data: $z_{\text{da}} = 1.12 \pm 0.0415$
	\item For the combination of datasets: $z_{\text{da}} = 1.1 \pm 0.0869$
\end{itemize}
Figure 7 indicate the comparison the evolution of deceleration parameter q with respect to redshift z  between $f(R)$ gravity and Rastall gravity. 
 \begin{figure*}
 	\centering
 	\includegraphics[width=14.5 cm]{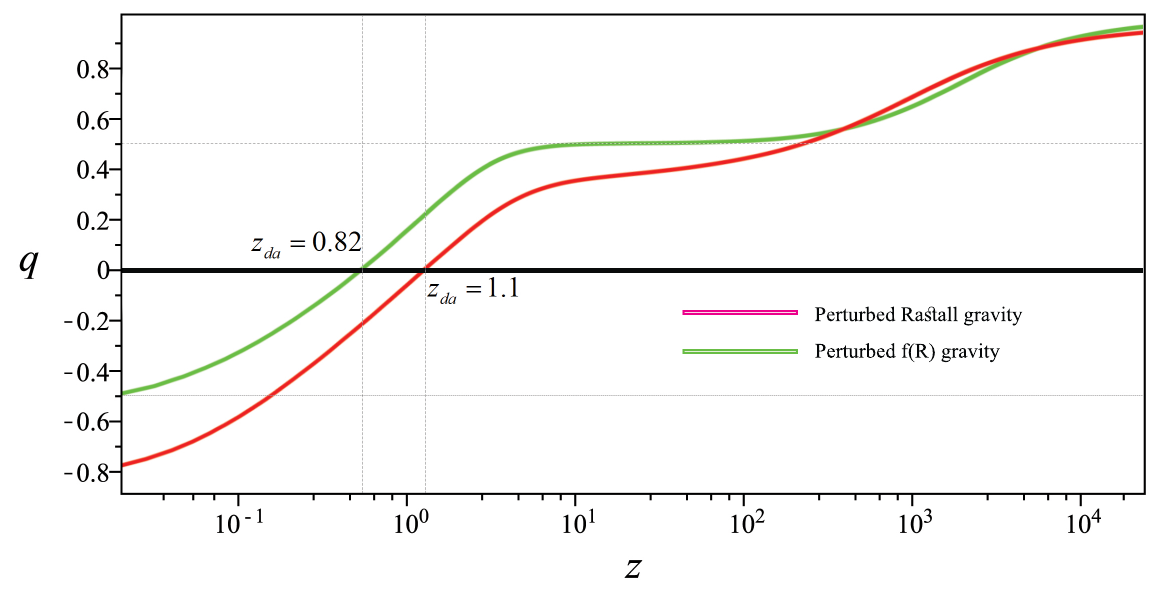}
 	\vspace{-0.42cm}
 	\caption{\small{Comparison of the deceleration parameter q(z) as a function of redshift z for different combinations of observational datasets. The analysis is performed within the framework of $f(R)$ gravity and Rastall gravity, highlighting the differences in the predicted  $z_{da}$ across these modified gravity models.}}\label{fig:omegam2}
 \end{figure*}

 We can concluded that in the Early universe, both model (Rastall gravity and $f(r)$ gravity ) can satisfy the circumstance of structure formation but in Late universe $f(r)$ gravity model can better explain the recent acceleration expansion of the universe. Figure 5 demonstrate the evolution of the universe from radiation dominated to dark energy dominated.  
    Also, all results that we mentioned in the above are in table 2,3.\color{black}
  \begin{figure*}
 	\centering
 	\includegraphics[width=14.5 cm]{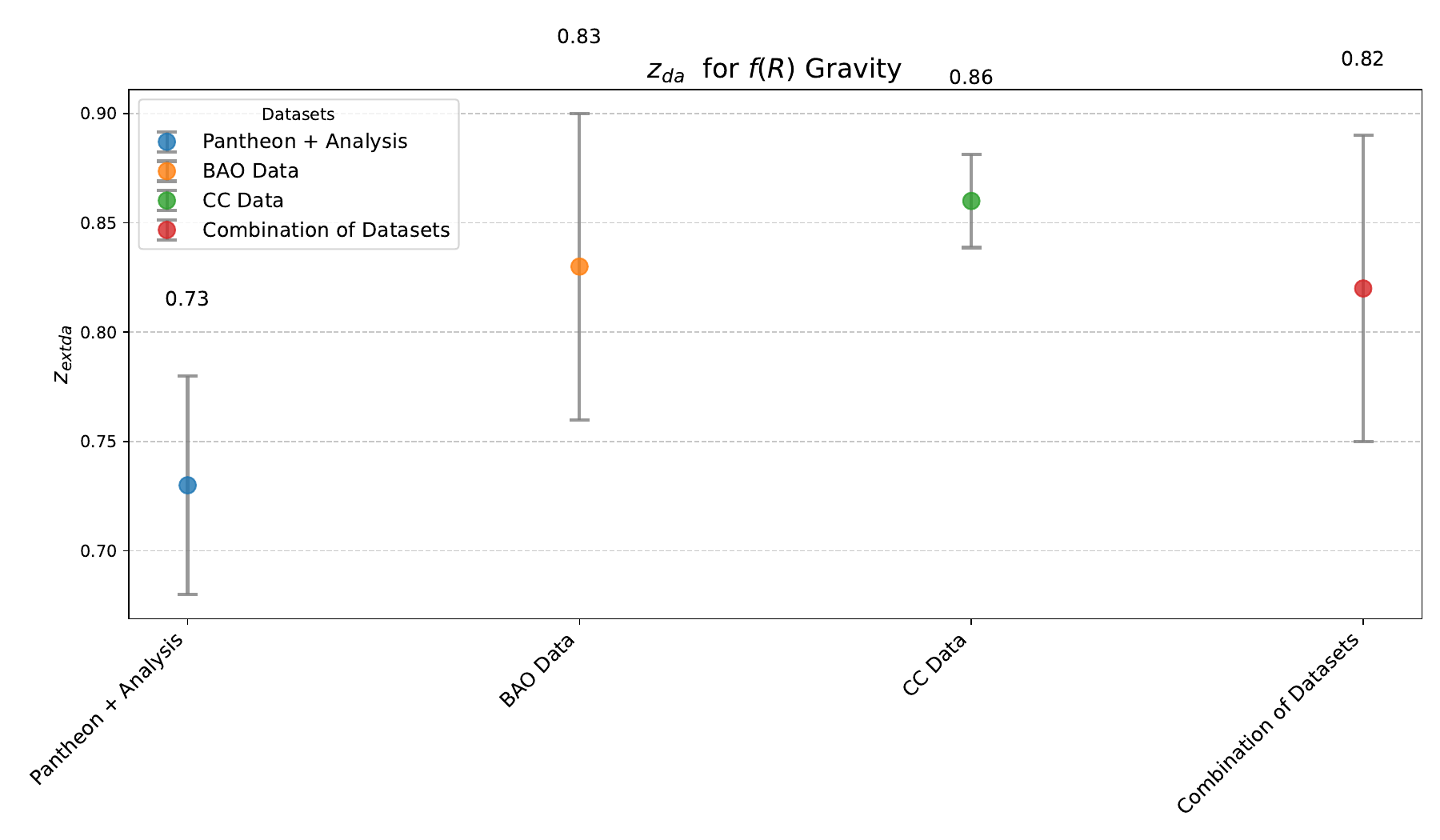}
 	\vspace{-0.42cm}
 	\caption{\small{Comparison the $z_{da}$ for $f(R)$ gravity for different combination of datasets.  }}\label{fig:omegam2}
 \end{figure*}
  \begin{figure*}
 	\centering
 	\includegraphics[width=14.5 cm]{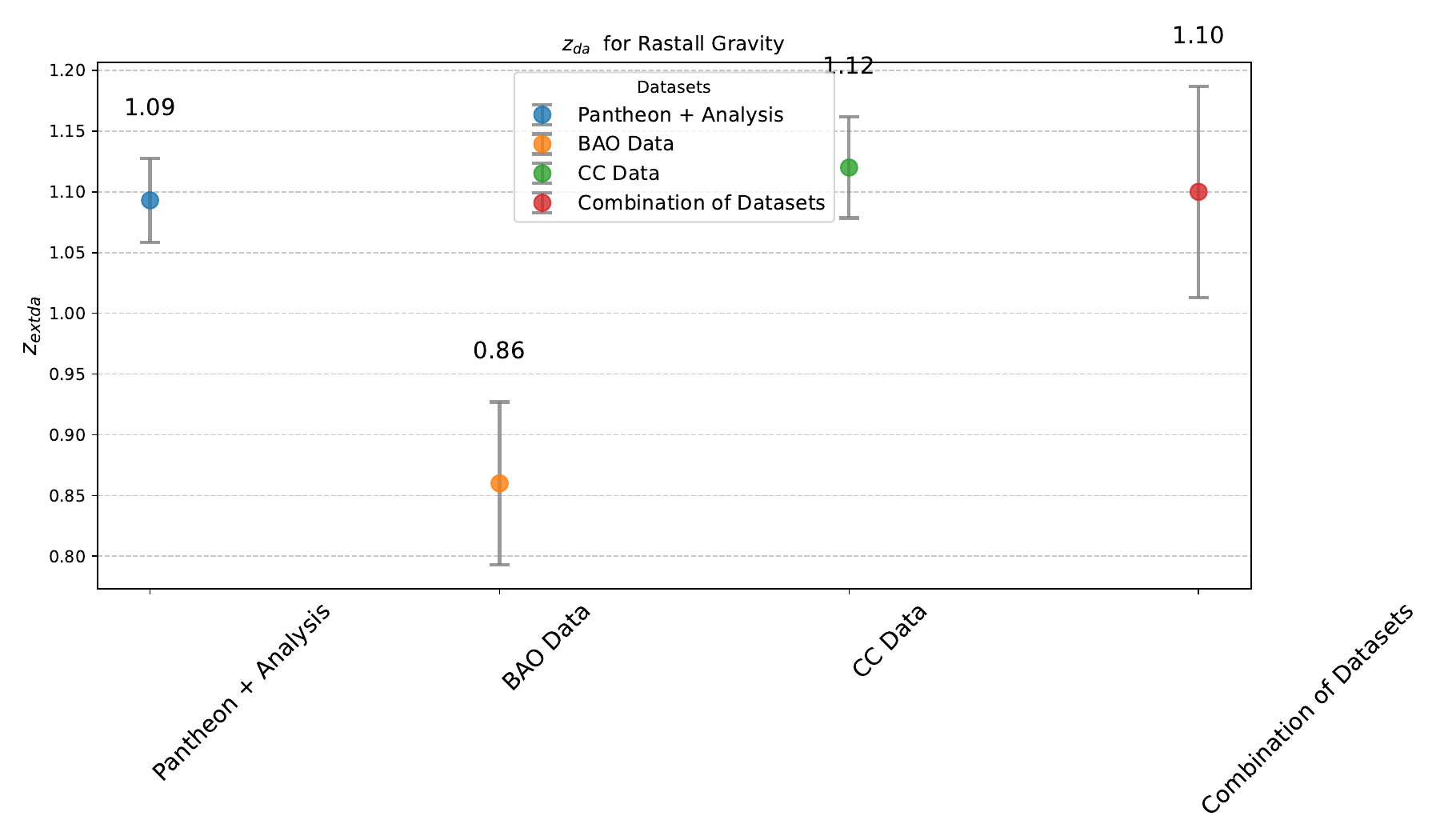}
 	\vspace{-0.42cm}
 	\caption{\small{Comparison the $z_{da}$ for Rastall gravity for different combination of datasets. }}\label{fig:omegam2}
 \end{figure*}

	\begin{table}
 	\centering
 	\caption{Summary of results for $f(R)$ gravity model.}
 	\begin{tabular}{|l|c|c|c|}
 		\hline
 		\textbf{Data Combination} & \textbf{Sound Speed} $c_s$ & \textbf{Jeans Wavenumber} $k_{\mathrm{J}}$ (Mpc$^{-1}$ c) & \textbf{DA Redshift Transition} $z_{\text{da}}$ \\ \hline
 		CMB + Pantheon + Analysis & $0.044 \pm 0.05$ & $0.000243$ & $0.73 \pm 0.05$ \\ \hline
 		CMB + CC Data & $0.056 \pm 0.041$ & $0.000267$ & $0.86 \pm 0.0213$ \\ \hline
 		CMB + BAO & $0.038 \pm 0.068$ & $0.000248$ & $0.83 \pm 0.07$ \\ \hline
 		CMB + BAO + Pantheon + Analysis + CC & $0.025 \pm 0.043$ & $0.000245$ & $0.82 \pm 0.07$ \\ \hline
 	\end{tabular}
 	\label{tab:fr_gravity_results}
	\end{table}

	\begin{table}
	\centering
	\caption{Summary of results for Rastall gravity model.}
	\begin{tabular}{|l|c|c|c|}
		\hline
		\textbf{Data Combination} & \textbf{Sound Speed} $c_s$ & \textbf{Jeans Wavenumber} $k_{\mathrm{J}}$ (Mpc$^{-1}$ c) & \textbf{DA Redshift Transition} $z_{\text{da}}$ \\ \hline
		CMB + Pantheon + Analysis & $0.077 \pm 0.0615$ & $0.000298$ & $1.093 \pm 0.0346$ \\ \hline
		CMB + CC Data & $0.069 \pm 0.073$ & $0.000301$ & $1.12 \pm 0.0415$ \\ \hline
		CMB + BAO & $0.046 \pm 0.05$ & $0.000274$ & $0.86 \pm 0.067$ \\ \hline
		CMB + BAO + Pantheon + Analysis + CC & $0.063 \pm 0.02$ & $0.000255$ & $1.1 \pm 0.0869$ \\ \hline
	\end{tabular}
	\label{tab:rastall_gravity_results}
	\end{table}

 \section{conclusion}

Our rigorous analysis consistently underscores the efficacy of both Rastall gravity and $f(R)$ gravity models in addressing various cosmic phenomena. However, when examining the late-time accelerated expansion of the universe, the $f(R)$ gravity model emerges as a more robust framework.

\textbf{For $f(R)$ gravity:} The combination of all datasets yields a sound speed of $c_{s} = 0.025 \pm 0.043$ and a Jeans wavenumber of $k_{\mathrm{J}} = 0.000245 \, \text{Mpc}^{-1} \, c$. The Deceleration-Acceleration (DA) redshift transition for $f(R)$ gravity is found to be $z_{\text{da}} = 0.82 \pm 0.07$, closely aligning with results from the $\Lambda$CDM model.

\textbf{For Rastall gravity:} The combination of all datasets constrains the sound speed to $c_{s} = 0.063 \pm 0.02$ and the Jeans wavenumber to $k_{\mathrm{J}} = 0.000255 \, \text{Mpc}^{-1} \, c$. The DA redshift transition is $z_{\text{da}} = 1.1 \pm 0.0869$, indicating a slightly higher transition redshift compared to $f(R)$ gravity.

\textbf{For the $\gamma$ parameter in Rastall gravity:} The combined analysis of all datasets constrains $\gamma$ to be $\gamma = 2.1 \pm 1.5$. This result suggests a moderate deviation from the standard scenario, with an uncertainty that reflects the complex nature of Rastall gravity's modifications to the conservation law.

Overall, the inherent flexibility of $f(R)$ gravity positions it as a more compelling alternative to the $\Lambda$CDM model in explaining the observed cosmic acceleration. In contrast, while Rastall gravity offers interesting theoretical prospects, its predictions for late-time acceleration present some challenges when compared to $f(R)$ gravity. This investigation not only enhances our understanding of the comparative strengths and limitations of these two models across different cosmic epochs but also provides a solid foundation for future research in cosmology, steering the exploration of cosmic evolution and its underlying dynamics.

\appendix
\section{Step-by-Step Derivation: From Christoffel Symbols to the Perturbed Metric}

Let's break down the process of moving from the calculation of Christoffel symbols to deriving the perturbed FRW metric.

\subsection{Compute the Christoffel Symbols (Step 1)}

The Christoffel symbols, which describe the connection coefficients of the metric, are calculated using the following formula:

\begin{equation}
\Gamma^\lambda_{\mu\nu} = \frac{1}{2} g^{\lambda\sigma} \left( \partial_\mu g_{\sigma\nu} + \partial_\nu g_{\sigma\mu} - \partial_\sigma g_{\mu\nu} \right)
\end{equation}

Here, \( g_{\mu\nu} \) is the perturbed metric that includes the scalar potentials \( \phi \) and \( \psi \). The perturbed metric is given by:

\begin{equation}
g_{00} = a^2(\eta)(1 + 2\phi), \quad g_{ij} = -a^2(\eta)(1 - 2\psi)\delta_{ij}
\end{equation}

We will now compute the Christoffel symbols step by step.

\subsubsection{Example: Compute \(\Gamma^0_{00}\)}

First, let's compute \( \Gamma^0_{00} \). This component involves the time-time part of the metric \( g_{00} \):

\begin{equation}
\Gamma^0_{00} = \frac{1}{2} g^{00} \partial_\eta g_{00}
\end{equation}

To find this, we need the inverse of \( g_{00} \), which is:

\begin{equation}
g^{00} = \frac{1}{a^2(\eta)(1 + 2\phi)}
\end{equation}

Next, calculate the partial derivative of \( g_{00} \) with respect to \(\eta\):

\begin{equation}
\partial_\eta g_{00} = \partial_\eta \left( a^2(\eta)(1 + 2\phi) \right) = 2a(\eta)\dot{a}(\eta)(1 + 2\phi) + a^2(\eta)\frac{\partial \phi}{\partial \eta}
\end{equation}

Now, substitute this into the expression for \( \Gamma^0_{00} \):

\begin{equation}
\Gamma^0_{00} = \frac{1}{2} \frac{1}{a^2(\eta)(1 + 2\phi)} \left( 2a(\eta)\dot{a}(\eta)(1 + 2\phi) + a^2(\eta)\frac{\partial \phi}{\partial \eta} \right)
\end{equation}

Simplifying this expression:

\begin{equation}
\Gamma^0_{00} = \frac{\dot{a}(\eta)}{a(\eta)} + \frac{\partial \phi}{\partial \eta}
\end{equation}

\subsubsection{Compute Other Christoffel Symbols}

Similarly, other components like \( \Gamma^0_{0i} \), \( \Gamma^i_{00} \), \( \Gamma^i_{j0} \), and \( \Gamma^i_{jk} \) can be computed. For example:

\begin{itemize}
	\item For \( \Gamma^0_{0i} \):
	
	\begin{equation}
	\Gamma^0_{0i} = \frac{\partial \phi}{\partial x^i}
	\end{equation}
	
	\item For \( \Gamma^i_{00} \):
	
	\begin{equation}
	\Gamma^i_{00} = -\frac{1}{a^2(\eta)}\frac{\partial \phi}{\partial x^i}
	\end{equation}
	
	\item For \( \Gamma^i_{j0} \):
	
	\begin{equation}
	\Gamma^i_{j0} = \frac{\dot{a}(\eta)}{a(\eta)}\delta^i_j + \frac{\partial \psi}{\partial x^j}
	\end{equation}
\end{itemize}

\subsection{Compute the Ricci Tensor (Step 2)}

The Ricci tensor \( R_{\mu\nu} \) is calculated using the Christoffel symbols:

\begin{equation}
R_{\mu\nu} = \partial_\lambda \Gamma^\lambda_{\mu\nu} - \partial_\nu \Gamma^\lambda_{\mu\lambda} + \Gamma^\lambda_{\lambda\rho} \Gamma^\rho_{\mu\nu} - \Gamma^\lambda_{\mu\rho} \Gamma^\rho_{\nu\lambda}
\end{equation}

Let's compute some components of the Ricci tensor:

\subsubsection{Compute \( R_{00} \)}

For \( R_{00} \):

\begin{equation}
R_{00} = \partial_\eta \Gamma^0_{00} - \sum_i \partial_i \Gamma^i_{00} + \Gamma^0_{00}\Gamma^0_{00} - \sum_i \Gamma^i_{00}\Gamma^0_{i}
\end{equation}

Substitute the previously calculated Christoffel symbols:

\begin{equation}
R_{00} = \partial_\eta \left( \frac{\dot{a}}{a} + \frac{\partial \phi}{\partial \eta} \right) - \sum_i \partial_i \left( -\frac{1}{a^2}\frac{\partial \phi}{\partial x^i} \right)
\end{equation}

Simplifying further:

\begin{equation}
R_{00} = \frac{\ddot{a}}{a} + \frac{\partial^2 \phi}{\partial \eta^2} - \frac{\nabla^2 \phi}{a^2}
\end{equation}

\subsubsection{Compute \( R_{ij} \)}

For the spatial components \( R_{ij} \):

\begin{equation}
R_{ij} = \partial_\eta \Gamma^0_{ij} - \partial_i \Gamma^0_{j} - \partial_j \Gamma^0_{i} + \Gamma^0_{00}\Gamma^0_{ij} - \Gamma^k_{i0}\Gamma^0_{jk} + \text{other terms}
\end{equation}

The calculation for \( R_{ij} \) is more involved but follows a similar substitution and simplification process.

\subsection{Einstein Tensor and Perturbed Einstein Equations}

Once the Ricci tensor \( R_{\mu\nu} \) is calculated, the Einstein tensor \( G_{\mu \nu} \) is found using:

\begin{equation}
G_{\mu\nu} = R_{\mu \nu} - \frac{1}{2} g_{\mu \nu} R
\end{equation}

Where \( R = g^{\mu\nu} R_{\mu \nu} \) is the Ricci scalar. The perturbed Einstein equations then relate \( G_{\mu \nu} \) to the stress-energy tensor \( T_{\mu \nu} \) via:

\begin{equation}
G_{\mu \nu} = 8 \pi G T_{\mu \nu}
\end{equation}

This step involves solving for the perturbations \( \phi \) and \( \psi \) by equating the components of \( G_{\mu \nu} \) to the corresponding components of \( T_{\mu \nu} \).

\subsection{Conclusion: Deriving the Perturbed Metric}

Through these steps, we've detailed how to go from the perturbed metric to the Christoffel symbols, then to the Ricci tensor, and finally to the Einstein tensor. The perturbed Einstein equations govern the dynamics of the scalar perturbations \( \phi \) and \( \psi \), leading to equations like the Poisson equation, which links these perturbations to matter density.

The final perturbed metric is:
\begin{equation}
	ds^2 = a^2(\eta) \left[ (1 + 2\phi)d\eta^2 - (1 - 2\psi) \delta_{ij} dx^i dx^j \right]
\end{equation}

\end{document}